\documentclass[a4paper]{article}\usepackage[]{graphicx}\usepackage[]{color}
\makeatletter
\def\maxwidth{ %
  \ifdim\Gin@nat@width>\linewidth
    \linewidth
  \else
    \Gin@nat@width
  \fi
}
\makeatother

\definecolor{fgcolor}{rgb}{0.345, 0.345, 0.345}

\usepackage{framed}
\makeatletter
 {\par\unskip\endMakeFramed%
 \at@end@of@kframe}
\makeatother

\definecolor{shadecolor}{rgb}{.97, .97, .97}
\definecolor{messagecolor}{rgb}{0, 0, 0}
\definecolor{warningcolor}{rgb}{1, 0, 1}
\definecolor{errorcolor}{rgb}{1, 0, 0}

\usepackage{alltt}

\usepackage[english]{babel}
\usepackage[utf8x]{inputenc}
\usepackage[T1]{fontenc}
\usepackage{tikz}
\usepackage{siunitx}
\usepackage{booktabs}
\usepackage{amsmath,amsfonts,amssymb,amsthm}
\usepackage{xspace}
\let\proglang=\textsf
\newcommand{\pkg}[1]{{\fontseries{b}\selectfont #1}}
\usepackage[a4paper,top=3cm,bottom=2cm,left=3cm,right=3cm,marginparwidth=1.75cm]{geometry}
\usepackage{amsmath}
\usepackage{graphicx}
\usepackage[colorinlistoftodos]{todonotes}
\usepackage[colorlinks=true, allcolors=blue]{hyperref}
\usepackage{natbib}
\usepackage{hyperref}
\usepackage{authblk}

\title{Parametric uncertainty in complex environmental models: a cheap emulation approach for models with high-dimensional output.}

 \author[1]{B. Swallow\thanks{ben.swallow@bristol.ac.uk} }
 \author[1]{M. Rigby}
 \author[2]{J.C. Rougier}
 \author[3]{A.J. Manning}
 \author[1]{M. Lunt} 
 \author[1]{S.  O'Doherty}
 \affil[1]{School of Chemistry, University of Bristol, UK}
 \affil[2]{School of Mathematics, University of Bristol, UK}
 \affil[3]{Met Office, Exeter, UK}
\date{}
\IfFileExists{upquote.sty}{\usepackage{upquote}}{}
\begin{document}
\maketitle

\begin{abstract}
In order to understand underlying processes governing environmental and physical processes, and predict future outcomes, a complex computer model is frequently required to simulate these dynamics.  However there is inevitably uncertainty related to the exact parametric form or the values of such parameters to be used when developing these simulators, with \emph{ranges} of plausible values prevalent in the literature.  Systematic errors introduced by failing to account for these uncertainties have the potential to have a large effect on resulting estimates in unknown quantities of interest.  Due to the complexity of these types of models, it is often unfeasible to run large numbers of training runs that are usually required for full statistical emulators of the environmental processes.  We therefore present a method for accounting for uncertainties in complex environmental simulators without the need for very large numbers of training runs and illustrate the method through an application to the Met Office's atmospheric transport model NAME.  We conclude that there are two principle parameters that are linked with variability in NAME outputs, namely the free tropospheric turbulence parameter and particle release height.  Our results suggest the former should be significantly larger than is currently implemented as a default in NAME, whilst changes in the latter most likely stem from inconsistencies between the model specified ground height at the observation locations and the true height at this location.  Estimated discrepancies from independent data are consistent with the discrepancy between modelled and true ground height.
\end{abstract}

\section{Introduction}
Physical and environmental models that replicate complex processes are frequently used to study and understand past, present and future events.  These computer models, or `simulators', replicate often complex processes that need to be accounted for when full experiments are unable to be run.  Sometimes the output of these simulators themselves are the quantities that scientists are interested in, but often these outputs are fed into other analyses and are merely an intermediate stage in the overall modelling process.  It is therefore vital that we understand and account for any uncertainties in these simulators if we are to avoid systematic errors and biases being introduced and propagated throughout the study.  Due to the complexity of the processes being modelled, even the computational power of modern computers is frequently not sufficient to allow these uncertainties to be studied fully. The ability to reduce the complexity of the simulator in order to explore uncertainties and sensitivities in it, is therefore of much interest.

Over the past few years there has been a dramatic increase in the use of statistical emulators to study and predict output from computer models.  The methods have gone from focusing on Gaussian process emulation of scalar outputs \citep[e.g.][]{santner03}, to more complex vector outputs \citep{rougier08, zhang15, overstall16}.  However, the output of computer models is often of a much higher dimension, and the methods required become increasingly more complex.  Some work has already been done on the emulation of output of large spatial models using thin-plate splines \citep{bowman16}.  However, not all high-dimensional outputs are of a spatial nature and alternative methods may require alternative representations.

Therefore, we propose a method that can take into consideration the uncertainties in the parameterisations of these simulators, using information based on expert elicitation, and incorporate this into an existing Bayesian hierarchical modelling framework to test what observed effects this additional uncertainty has on the overall estimates of interest.  

This is a full report accounting the methodology developed for and applied to the Met Office's atmospheric dispersion model (NAME).  The methodology is applied to a single month of observations, specifically January 2013, from four observation stations across the UK and Ireland, in order to generate new estimates of UK emissions of methane and their associated uncertainties.

\citet{harvey16} constructed a Bayes linear emulator for NAME to be used in predicting natural hazards, specifically predicting volcanic ash concentrations and transport.  However, many of the parameters and the nature of the model output in this instance is very different from that used in greenhouse gas flux inversions.

The report is outlined as follows. In Section \ref{sec:app} we outline the application which will be used to illustrate the methods.  The following sections will then introduce the main steps of the emulation process, specifically referring back to the example outlined in Section \ref{sec:app}.  Whilst the methods are tuned specifically to this particular example, they are widely applicable to a range of different model inputs and outputs.  The section headings for Sections \ref{sec:pars} to \ref{sec:inversion} relate to the general requirement of that aspect of the emulation process, but the specifics within each section can easily be modified to fit the nature of the problem at hand.  These relate to the specification of the input parameters and subsequent model training, reduction of the model output, determination of the statistical relationship between input and output and the incorporation of these parameters into the overall inversion setup.

Due to the complexity of the methods developed and the range of different modelling approaches utilised, it was necessary to conduct the analyses in a range of programming languages.  The pre-processing is conducted in \proglang{Python}, the statistical model fitting in Section \ref{sec:statmod} in \proglang{R} and the computationally intensive parameter estimation in \proglang{Fortran 90}, extended from code developed from \citet{lunt16}.

\section{Application}\label{sec:app}

In order to illustrate the methodology outlined in this manuscript, we will use a specific application in atmospheric chemistry.  The overall aim is to estimate fluxes of greenhouse gases, in particular methane (CH$_4$), from mole fractions observations from a network of observation stations across the UK and Ireland.  We use the mole fraction observations to produce top-down estimates of emissions of methane using a chemical transport model to link the two in an inversion framework.  The inversion is a hierarchical Bayesian inversion where all forms of uncertainty in the parameters are estimated directly from the data.  Initial prior beliefs are updated through the data to provide a posterior estimate of the parameters of interest.  These have been shown to improve estimates of emissions and their relative uncertainties compared to approaches where uncertainties are assumed fixed and known \citep{ganesan14}.  The parameters are estimated in a Markov chain Monte Carlo (MCMC) framework \citep[e.g.][]{gelman13}.  These will be now be discussed in more detail.

\subsection{The data}
\begin{figure}
\centering
\includegraphics[height=0.3\textheight]{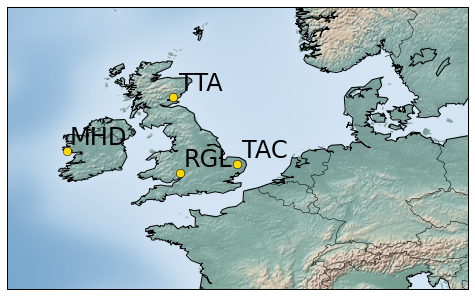}
\includegraphics[height=0.3\textheight]{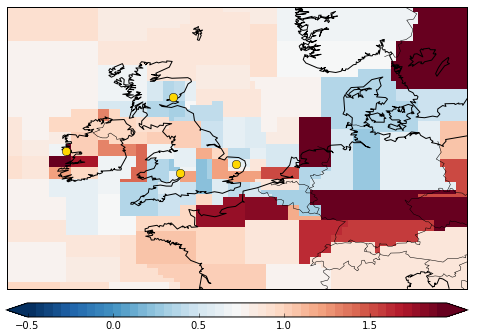} 
\caption{Location of observation sites (top) and an example prior flux scaling map showing the fixed regions for which fluxes are estimated (bottom).}
\label{fig:regions}
\end{figure}

The data consist of six-hourly averaged mole fractions obtained from four sites across the UK and Ireland for January 2013.  The sites and their locations are outlined in Table \ref{tab:sites} and shown in Figure \ref{fig:regions}.  The observations are subject to missing data for various reasons, and the total number of observations for each site are also given in Table \ref{tab:sites}.

\begin{table}[!ht]
\centering
\begin{tabular}{lcccccc}
\toprule
Site name & Abbreviation & Site no. & Latitude & Longitude & Inlet height (magl) & No. observations\\
\midrule
Mace Head & MHD & 1 & 53.33 & -9.90 & 10 & 120 \\
Ridge Hill & RGL & 2 & 51.99 & -2.54 & 90 & 114 \\
Tacolneston & TAC & 3 & 52.52 & 1.14 & 100 & 119 \\
Angus & TTA & 4 & 56.55 & -2.99 & 222 & 67 \\
\bottomrule
\end{tabular}
\caption{Locations and details pertaining to the four observation sites used in the inversion.}
\label{tab:sites}
\end{table}

\subsection{The simulator: NAME}
\begin{figure}[!ht]
\centering
\includegraphics[width=0.9\textwidth,trim={0 3cm 0 3cm},clip]{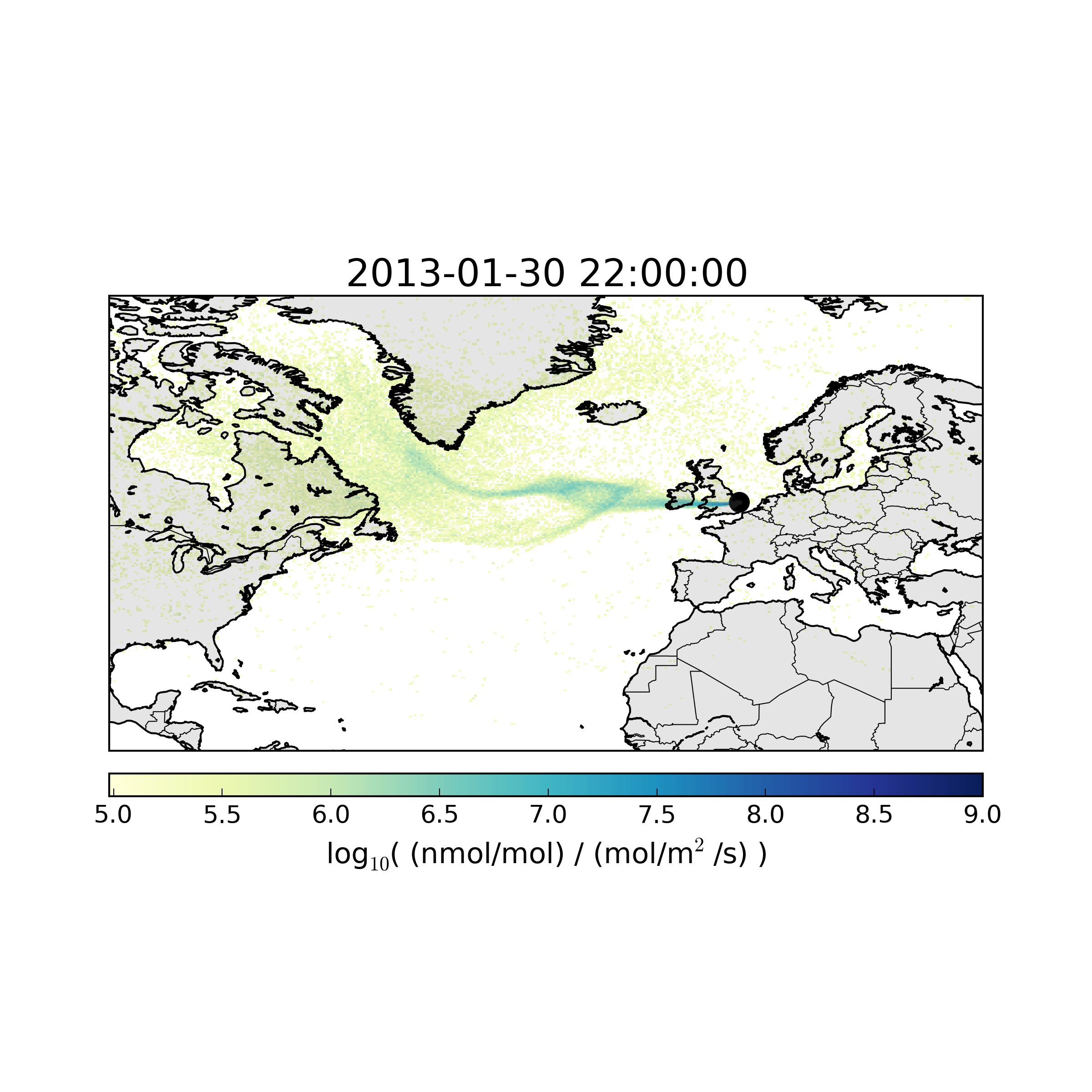} 
\caption{A two-hourly footprint for Tacolneston generated from NAME with the default parameters.}
\label{fig:fp}
\end{figure}

The Met Office's Numerical Atmospheric-dispersion Modelling Environment (NAME) \citep{Jones2007} is a Lagrangian particle trajectory model, which replicates the transport of emissions from sources by following released `particles' through a model atmosphere, driven by meteorological fields.  For use in greenhouse gas emission inversion estimations, NAME is run backwards in time, such that theoretical particles are released from each of the monitoring sites and followed backwards in time, replicating the process of gases being transported through the atmosphere from sources at the surface to the observation stations at which they are detected.  In particular, the model determines when and where particles released from the monitoring station spend time in the lower 40m of the modelled atmosphere, which is represented through a user-defined resolution topography (in this case 25km resolution to link in with meteorology data used).  A map of where the particles have concentrated within the lower 40m, called a 'footprint', can then be plotted to show regions that each observation was sensitive to.  Figure \ref{fig:fp} shows one such footprint for a two hour period at a monitoring site at Tacolneston, UK.  The dark areas show the places at which the particles released into the computational domain have concentrated with the lower atmosphere, indicating that the monitoring site would be sensitive to theoretical emissions from these regions over this two hour period.  

The model accounts for numerous complex environmental and atmospheric phenomena through mathematical parameterisations.  Single parametric formulations, often with fixed coefficients, must be given to these complex and often changing processes.  If these parameterisations are incorrect, then systematic errors will be implemented into the inverse modelling process; similarly estimates of uncertainty could be severely underestimated.  Therefore it is important to account for uncertainties in the formulation of this model and the parameters that are fed into it.

\subsection{Sensitivity matrix \textit{H}}
\begin{figure}
\centering
  \begin{tikzpicture}
  \node (img)  {\includegraphics[height=0.4\textheight]{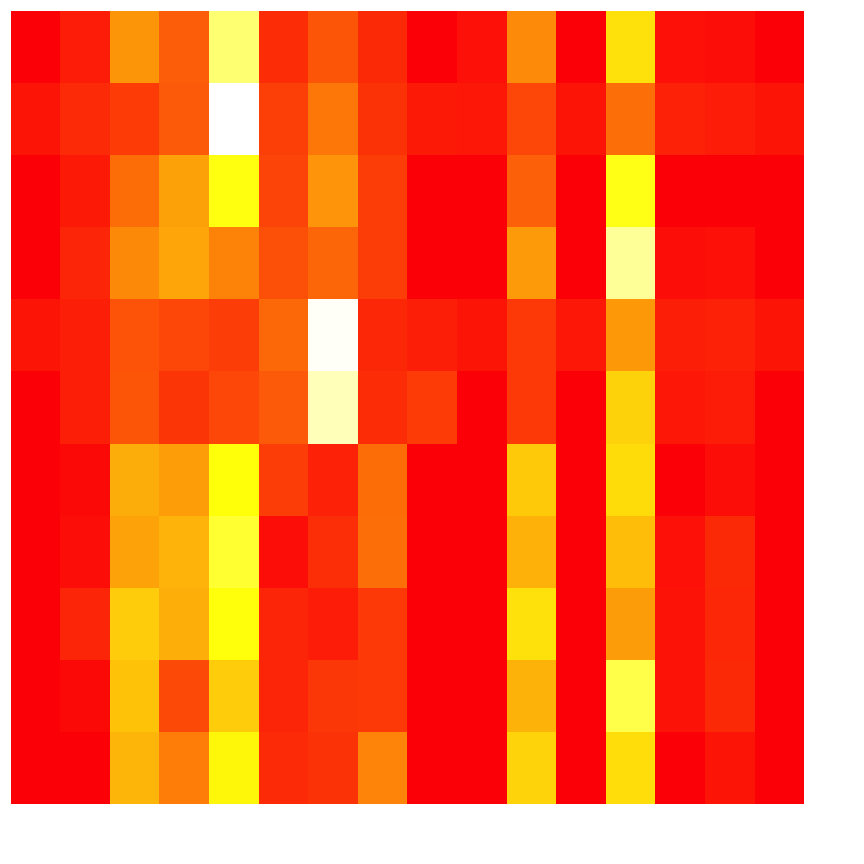}};
  \node [above=of img,node distance=0cm, scale=0.6, yshift=-1.5cm] {\includegraphics[scale=0.25]{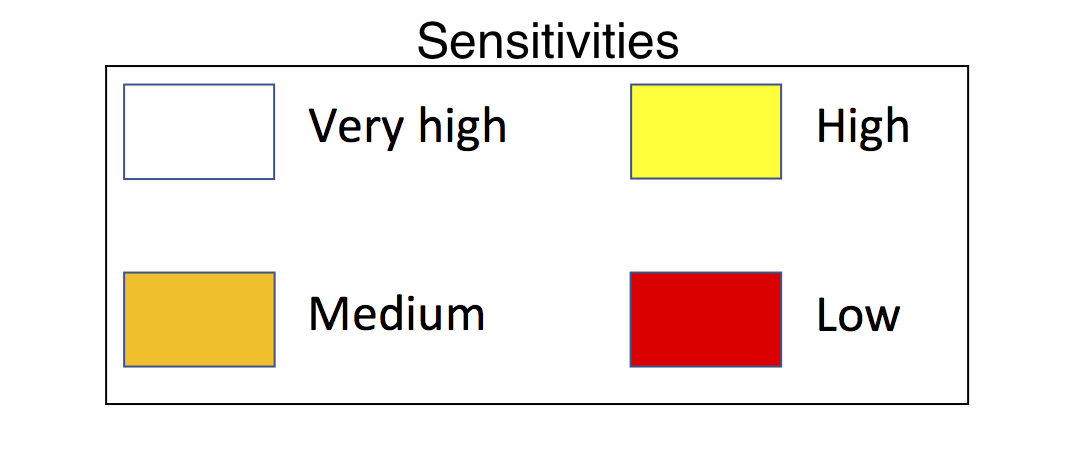}};
  \node[below=of img, node distance=0cm, yshift=1cm] {Regions};
  \node[left=of img, node distance=0cm, rotate=90, anchor=center,yshift=-0.7cm] {Observations};
 \end{tikzpicture}
\caption{Heatmap of a section of $H$ matrix, showing sensitivity of observations to changes in emissions from a given region.   The smooth colour scale from red (low) to white (high) shows sensitivity of observations to emissions from each region.}
\label{fig:heatmap}
\end{figure}

Once the footprint has been generated for each site as described in the previous section, a matrix $H$ representing the sensitivity of each observation, for a given time period and site, to changes in emissions from each region, is outputted.  Figure \ref{fig:heatmap} shows a graphical representation of a section of one such $H$ matrix.  Dark coloured squares relate to low sensitivity of that observation to the emissions from the corresponding regions, whilst light regions reflect high sensitivity.

In the general inverse model framework, we attempt to estimate a vector $\textbf{x}$ of trace gas fluxes using mole fraction observations $\mathbf{y}$.  The matrix $H$ is then a linear map accounting for various processes that relate the fluxes emitted from the surface to their subsequent measurements at the measurements sites, i.e.

\begin{equation}\label{eqn:inversion}
\mathbf{y}=H \mathbf{x} + \epsilon 
\end{equation}

\noindent where $\epsilon$ is usually assumed to be a zero-mean normally-distributed error, although other alternatives have also been used \citep{ZM15}.  It has previously been assumed that $H$ is fixed and known a-priori.

\section{Choosing the parameters and ranges}\label{sec:pars}
Consultations with the Met Office model developers highlighted 11 NAME input parameters of interest that were considered to be uncertain and have the potential to affect the output (footprint) generated.  Expert elicitation provided ranges of plausible values for these 11 parameters of interest (Table \ref{tab:parranges}).  Upon closer inspection of the code, it was apparent that there were some dependencies between the boundary layer turbulences and the free tropospheric turbulence, limiting the possible difference between the two turbulence schemes.  To allow for a reduction in this dependence, four additional parameters were added to the sensitivity analysis, corresponding to the limit on the boundary layer diffusivities as a proportion of the free tropospheric diffusivities. Limits were also placed on the variances and the Lagrangian timescales, which meant there would still be dependencies between the two schemes but these should be of a lower degree than previously.  These additional parameters brought the total to 15.

In NAME, the free tropospheric turbulence variance is multiplied by the Lagrangian timescales to calculate the eddy diffusivities $K$, that is:

\begin{equation}
K_{u,w}=\sigma_{u,v}^2\tau_{u,w}.\nonumber
\end{equation}

As it is only the diffusivity parameters $K$ that are ultimately the key turbulence parameters in the version of NAME used, we merely scale these parameters up and down, rather than the individual variances and Lagrangian timescales.  It was assumed that there would be a correlation between the free tropospheric turbulence parameters in the horizontal (u) and vertical (w) directions.  Therefore these were considered a single parameter for the purpose of this analysis.  The range of values for these parameters were calculated by initially generating plausible values for $\sigma_u$; these parameters were then converted into $K_u$ by squaring and multiplying by the Lagrangian timescales.  The scale of this new $K_u$ compared to the default $K_u$ was then calculated before this scaling was applied to the default $K_w$.  As plausible ranges for the $K$ had also been given, all $\sigma$s and $K$s were checked to ensure they fell within their respective ranges of plausible values before the analysis was run.  The Lagrangian timescales $\tau$ were kept constant and the inputted standard deviations were then equal to the square root of $\sqrt{K/\tau}$.  This both reduces the number of runs necessary and ensures that resulting changes in the footprints can be more easily allocated to the appropriate parameters.  Treating the two parameters independently would cause confounding in the estimation process, as the changes would be observed in the product of the variances and timescales, not their individual values.  For the boundary layer turbuelences, a scheme without velocity memory is employed to reduce the computational requirements of the model run.

\begin{table}[!ht]
\centering
\resizebox{\columnwidth}{!}{%
\begin{tabular}{lcccccc}
\toprule 
Parameter & Acronym & Notation & Type & Transformation & Default & Range \\
\midrule
MBL (m) & MBL & $\xi_1$ & Const & logit & 40* & [40,100] \\
Free tropospheric turbulence ($\sigma_u$/$\sigma_w$) & FTT & $\xi_2$ & Const & log & 0.25/0.1 & [0.06,0.82]/[0.02,0.35]\\
Unresolved mesoscale motions ($\sigma_u$) & UMM & $\xi_3$ & Const & log  & 0.8 & [0.16,0.85] \\
Boundary stable horizontal ($\sigma_u$) & BLHS & $\xi_4$ & Scale & logit & 1.0 & [0.8,1.8] \\
Boundary stable vertical ($\sigma_w$) & BLVS & $\xi_5$ & Scale & logit & 1.0 & [0.7,1.3] \\
Boundary unstable horizontal ($\sigma_u$) & BLHU & $\xi_6$ & Scale & logit & 1.0 & [0.8,1.8] \\
Boundary unstable vertical ($\sigma_w$) & BLVU & $\xi_7$ & Scale & logit & 1.0 & [0.7,1.3] \\
Boundary stable horizontal lim & LHS & $\xi_8$ & Scale & logit & 1.0 & [0.8,1.8] \\
Boundary stable vertical lim & LVS & $\xi_9$ & Scale & logit & 1.0 & [0.7,1.3] \\
Boundary unstable horizontal lim & LHU & $\xi_{10}$ & Scale & logit & 1.0 & [0.8,1.8] \\
Boundary unstable vertical lim & LVU & $\xi_{11}$ & Scale & logit & 1.0 & [0.7,1.3] \\
Boundary layer depth & BLD & $\xi_{12}$ & Scale & logit & 1.0 & [0.5,1.5] \\
Release height (m) & Z & $\kappa_{j,1}$ & Diff & logit & $z$ & [-2*z,2*z + 100] \\
Release latitude ($\si{\degree}$) & X & $\kappa_{j,2}$ & Diff & logit & $x$ & $x$ + [-0.05,0.05] \\
Release longitude ($\si{\degree}$) & Y & $\kappa_{j,3}$ & Diff & logit & $y$ & $y$ + [-0.05,0.05] \\
\bottomrule
\end{tabular}}
\caption{NAME simulator parameter details.  Parameter type, transformation used, default value and range of plausible values are given. For the `diff' parameters, $z$ is the height of the monitoring sensor and $x$ and $y$ the latitude and longitude of the monitoring site respectively. `Const' parameters are numerical values of physical quantities; scale parameters are proportional to the default value of the parameter or function. *MBL is a user-defined parameter.  The default value here is the number usually used to generate the fixed $H$ matrices for these applications.}
\label{tab:parranges}
\end{table}

Of the 15 parameters, $I=12$ were considered to be site-invariant, whilst the other $L=3$ (those relating to the particle release location in the x, y and z directions), were allowed to vary by site.  The two tropospheric turbulence parameters, $\sigma_u$ (horizontal) and $\sigma_w$ (vertical), were assumed to be correlated and hence were amalgamated into a single parameter (with the second parameter scaled up or down by an equivalent amount compared to the default value).  

The vector of input parameters $\mathbf{\theta}$ was therefore defined as a combination of site-invariant parameters $\xi_i$ and site-specific parameters $\kappa_{jk}$:

\begin{equation}
\mathbf{\theta}=\{\{ \xi_i \},\{ \kappa_{jl}\}\}, \hspace{10pt} \textnormal{for} \hspace{10pt} i=1,\cdots,I, \hspace{10pt}j=1,\cdots,n_{sites} \hspace{10pt}\textnormal{and}\hspace{10pt}l=1,\cdots,L\nonumber
\end{equation}

\section{Training the model}\label{sec:training}
\subsection{Generating parameter values}

In order to understand how changes in the input parameters affect the output of NAME, we need to run the simulator at many combinations of input parameters.  This will then enable us to `train' the statistical model developed in the following sections.  Many combinations of parameter values spanning the full ranges of plausible values from Table \ref{tab:parranges} need to be generated; this is most commonly accomplished using a Latin hypercube (LHC) design \citep{mckay79}.  Specifically, we use of a 24-dimensional maxi-min Latin hypercube with entries $p=1,...,50$ (see \citet{moon11} for discussion), maximising the parameter space covered by the model runs, whilst also allowing interactions between parameters to be accounted for.  This gives a matrix of 50 rows of a total of $12+3*n_{sites}$ parameters (columns).  The first 12 columns relating to the site-invariant parameters $\xi_i$ and the last 12 being the site-specific $\kappa_{jk}$.  The first few combinations of parameters are shown in Table \ref{tab:parvals}.  In addition to the 50 parameter combinations generated by the LHC, we already have the values of output from the default parameter values and this was included as an additional parameter combination in the analyses and set to $p=0$.

\begin{table}[ht]
\centering
\resizebox{\linewidth}{!}{\begin{tabular}{cccccccccccccccc}
  \hline
p & MBL & FTT & UMM & BLHS & BLVS & BLHU & BLVU & LHS & LVS & LHU & LVU & BLD & X$_1$ & Y$_1$ & Z$_1$ \\ 
  \hline
0 & 40.00 & 0.25 & 0.80 & 1.00 & 1.00 & 1.00 & 1.00 & 0.50 & 0.50 & 0.50 & 0.50 & 1.00 & -9.90 & 53.33 & 10.00 \\ 
1 & 94.93 & 0.25 & 0.75 & 1.20 & 0.66 & 1.32 & 1.40 & 0.41 & 1.06 & 1.07 & 0.56 & 0.85 & -9.95 & 53.35 & 33.04 \\ 
  2 & 96.53 & 0.13 & 0.59 & 1.17 & 1.42 & 1.25 & 1.37 & 0.90 & 0.64 & 0.67 & 0.42 & 1.05 & -9.92 & 53.29 & 107.57 \\ 
  3 & 92.53 & 0.04 & 0.68 & 1.09 & 0.83 & 2.27 & 0.76 & 0.58 & 0.70 & 0.65 & 0.86 & 1.34 & -9.87 & 53.38 & 28.52 \\ 
  4 & 68.22 & 0.25 & 0.67 & 1.06 & 1.31 & 1.34 & 1.67 & 0.65 & 0.87 & 0.11 & 0.43 & 0.76 & -9.91 & 53.29 & 0.25 \\ 
  5 & 78.46 & 0.26 & 0.35 & 1.58 & 0.58 & 0.90 & 1.53 & 0.43 & 0.33 & 0.36 & 0.79 & 0.70 & -9.91 & 53.33 & 66.65 \\ 
  6 & 95.10 & 0.29 & 0.57 & 1.82 & 1.21 & 1.26 & 0.65 & 0.70 & 0.67 & 0.95 & 0.16 & 1.14 & -9.89 & 53.29 & 75.11 \\ 
  7 & 83.45 & 0.16 & 0.16 & 1.53 & 1.64 & 0.69 & 0.86 & 0.86 & 0.61 & 0.94 & 0.52 & 0.79 & -9.86 & 53.30 & 16.42 \\ 
  8 & 50.43 & 0.32 & 0.71 & 2.19 & 1.37 & 1.68 & 0.70 & 1.01 & 1.09 & 0.34 & 0.88 & 1.08 & -9.88 & 53.28 & 55.52 \\ 
  9 & 85.28 & 0.21 & 0.83 & 1.51 & 1.61 & 0.83 & 0.74 & 0.30 & 0.95 & 1.05 & 0.66 & 1.36 & -9.92 & 53.31 & 25.33 \\ 
  10 & 90.53 & 0.31 & 0.66 & 2.08 & 0.86 & 1.02 & 1.63 & 0.14 & 0.75 & 0.15 & 0.20 & 1.29 & -9.86 & 53.36 & 108.85 \\ 
\vdots &  \vdots & \vdots & \vdots & \vdots & \vdots & \vdots & \vdots & \vdots & \vdots & \vdots & \vdots & \vdots & \vdots & \vdots & \vdots \\ 
   \hline
\end{tabular}}
\caption{Example parameter combinations for one of the four observation sites.} 
\label{tab:parvals}
\end{table}

It is advisable when constructing emulators to have an idea of unfeasible parameter combinations, or outputs that are non-physical, as these can be removed from the analysis.  In this application it is difficult to conceptualise parameter combinations or outputs that would be deemed unfeasible other than those outside the ranges obtained during expert elicitation.  Therefore, all input parameter combinations within the ranges proposed through the expert elicitation process, and their corresponding outputs, were considered feasible.

\subsection{Running the simulator}
\begin{figure}
\centering
\includegraphics[width=0.9\textwidth,trim={0 3cm 0 3cm},clip]{fp_TAC_fixed-0.png} 
\includegraphics[width=0.9\textwidth,trim={0 3cm 0 3cm},clip]{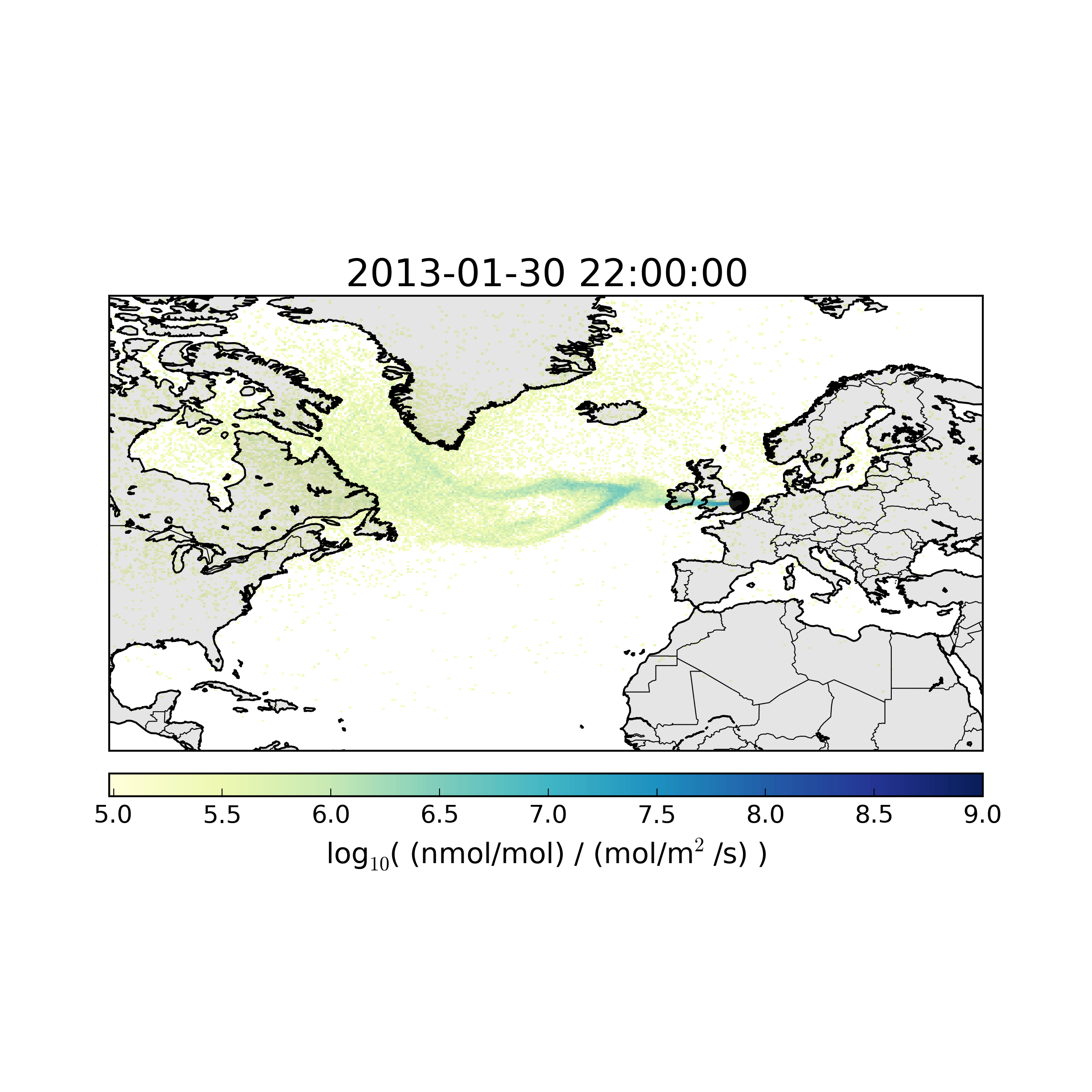}
\caption{Two equivalent two-hourly footprints for Tacolneston generated from NAME with different input parameters. Any differences between the two are purely down to the changes in input parameters.}
\label{fig:fp2}
\end{figure}

\begin{figure}
\centering
  \begin{tikzpicture}
  \node (img)  {\includegraphics[width=0.43\textwidth]{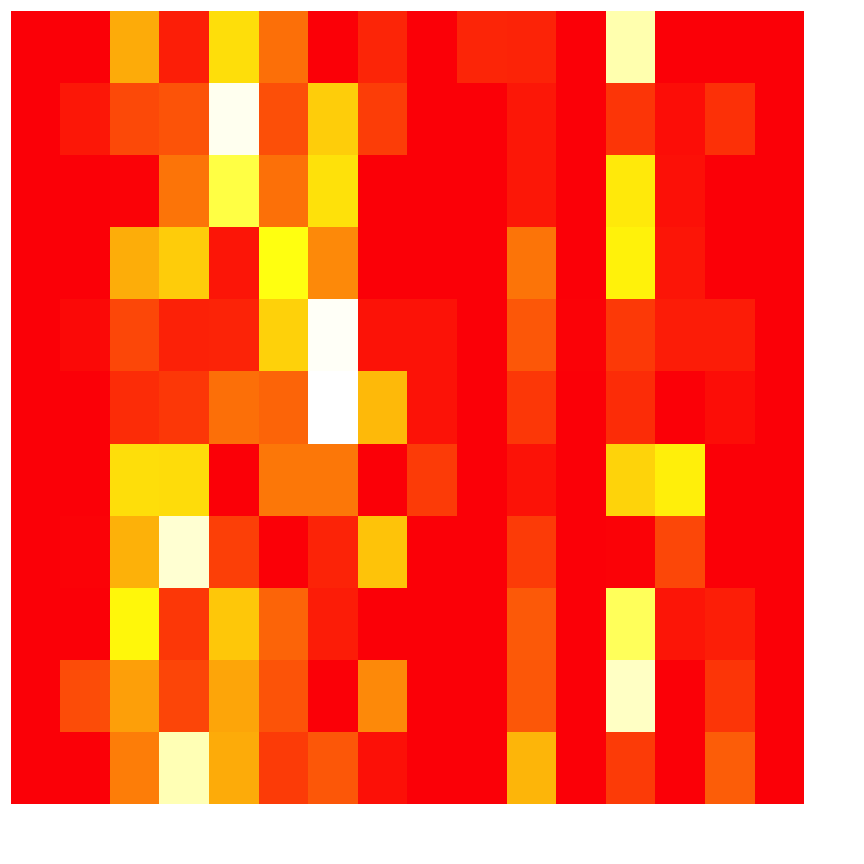}};
  \node [above right=of img,node distance=0cm, scale=0.6, yshift=-1.5cm, xshift=-5cm] {\includegraphics[scale=0.25]{"heatmap_key"}};
  \node[below right =of img, node distance=0cm, yshift=1cm, xshift = -1.5cm] {Regions};
  \node[left=of img, node distance=0cm, rotate=90, anchor=center,yshift=-0.7cm] {Observations};
  \node [right=of img, node distance=1cm, xshift=0cm] {\includegraphics[width=0.43\textwidth]{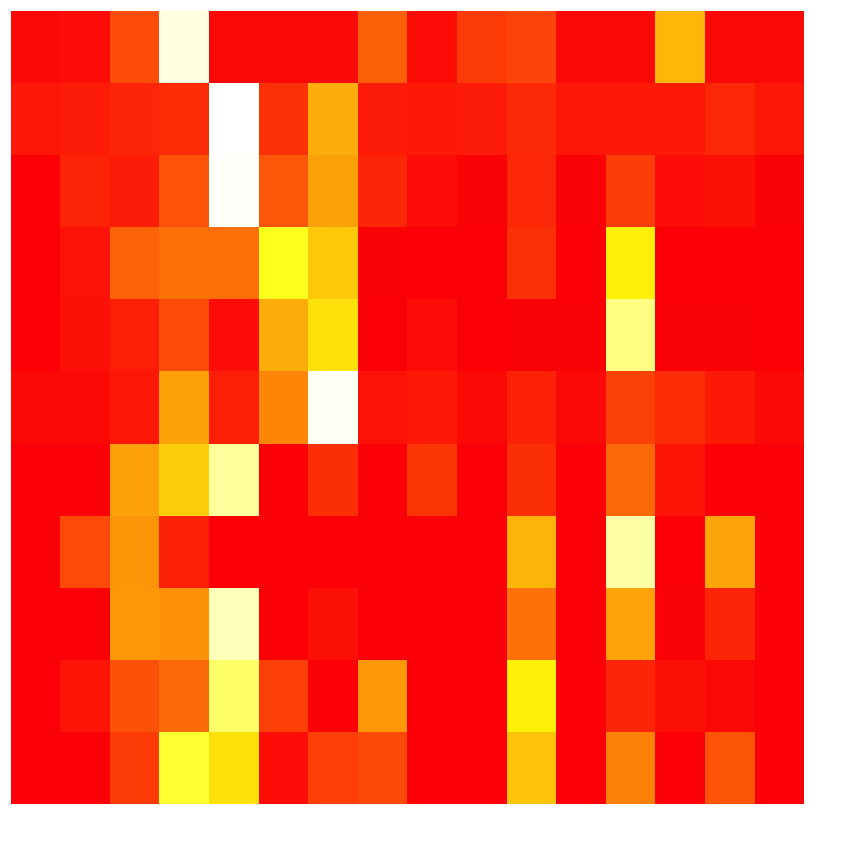}};
 \end{tikzpicture}
\caption{Two example heatmaps of the same section of $H$ as in Figure \ref{fig:heatmap} for different input parameter combinations. Once again, the smooth colour scale from red (low) to white (high) shows sensitivity of observations to emissions from each region.}
\label{fig:varheatmaps}
\end{figure}

The next stage is to run the simulator for each of the parameter combinations generated above.  In the case of NAME, this means replacing each of the default values for the parameters by their new values.  Some of these are simply input variables in the simulator script file, whilst others are parameters within the main model code and require a search and replace algorithm followed by recompilation of the main code in order that the changes are implemented.  Due to the computational demand of running all the simulations across sites and parameter combinations, high-performance computing services were used to run different simulations in parallel \footnote[1]{\url{http://www.bris.ac.uk/acrc/}}.  For each parameter combination, a full 30 day run in backwards mode was conducted for each site for the whole month of interest.  Two example footprints and their corresponding zoomed-in $H$ heatmaps for the same site and time period as in Figures \ref{fig:fp} and \ref{fig:heatmap} are shown in Figures \ref{fig:fp2} and \ref{fig:varheatmaps} respectively; the discrepancies between the three is purely down to the different parameters used in each case.

Each of the $p=0,\cdots,n$ simulator outputs (i.e. footprints) are converted to the sensitivity matrix $H$ for the chosen month.  A fixed grid developed for the Inversion Technique for Emission Modelling (InTEM) system is used, which splits up the European domain into 137 regions, with smaller regions, and hence greater sensitivity, around the monitoring locations \citep{manning12}.  In addition to these 137 regions, 12 additional larger regions are added around the borders of the inversion domain.  Each of the $H_p$ matrices are therefore of dimension $[n_{obs}, 149]$, corresponding to a single combination of parameter values denoted $\mathbf{\theta}_p$.

\section{Output dimension reduction}\label{sec:dimred}
Despite the fact that a full footprint is reduced to a single sensitivity matrix, which is considerably simpler than the full footprints, the number of individual elements to $H$ is still prohibitively large for a full Gaussian process emulator to be a viable option.  For the single month and four sites used in this study, $H$ consists of over 60,000 elements.  In order to make it realistic to model the uncertainty in $H$ as a function of the input parameters, it is first necessary to conduct dimension reduction on the full $H$ matrix such that we can then study uncertainty in the principal areas of variation before scaling this back up to the full $H$ matrix.  To do this, we use singular value decomposition (SVD).  SVD is a factorisation a matrix $X$ into the form $UDV^{\intercal}$, where $D$ is a diagonal matrix with non-negative elements on the diagonal.

At this point, the sections of $H$ corresponding to each site are treated as a separate entity that requires prediction.  That is, the rows corresponding to each site, of which there are the same number as the number of observations for that site, are separated out for the remainder of this section.  This step is taken as whilst it is reasonable to assume that the principal singular values corresponding to different sites are likely to reflect similar abstract concepts, the regions (or similarly the columns of $H$) that each site is most sensitive to are the ones closest to it.  These will differ by site and hence the patterns of reallocation of the singular values through $U$ and $V$ would be very different. Observation of the $U$ and $V$ matrices within and between sites suggested there was little difference in either across parameter ranges within a site, but differences were much greater between sites.  We also allow some of the parameters to be site specific, and these site-specific parameters are only allowed to impact the singular values relating to that site.

To calculate the SVD, we firstly sweep the row, column and overall mean of the default $H$ matrix, (that is the matrix generated by the default parameter values), into a new matrix $H_m$ before subtracting this from the $n+1$ different sensitivity matrices generated by NAME at each of the training parameter combinations. Each of these centred matrices (denoted $H^c$) is then decomposed using singular value decomposition into

\begin{equation}
H^c_{p,[s]}=U_p D_p V_p^{\intercal} \nonumber
\end{equation}

\noindent where $H^c_{.,[s]}$ denotes the rows of the centred $H$ corresponding to observations at site $s$, the $D_p$ are diagonal matrices of the singular values in which the singular values are ordered by magnitude.  Once we have calculated the singular value decomposition of each of the $n+1$ sensitivity matrices, we have a set of $n+1$ vectors of singular values $\mathbf{s_p}$ with each corresponding to a single vector of parameters $\mathbf{\theta_p}$.  

As there was generally very little variation in $U$ and $V$ within each site, we further simplify by only assuming that 

\begin{equation}
H^c_{0,[s]}=U_{0,s} D_0 V_{0,s}^{\intercal}, \nonumber
\end{equation}

\noindent and left and right multiply each addition $H^c$ by the default matrices to generate equivalent $D$ matrices.  

\begin{equation}
D_p=V_{0,s} H^c_{p,[s]} U_{0,s}^{\intercal},\hspace{2pt} \text{for }\hspace{2pt}p\neq 0 \nonumber
\end{equation}

At this point, in order to study uncertainty in $H$, we instead study uncertainty in $D$ or specifically in $\mathbf{s}$, the diagonal elements of $D$.

\section{Determining the statistical relationship between input and reduced output}\label{sec:statmod}
The overall aim of this method is to remove the need to further use the simulator for additional combinations of parameter inputs that have not already been run. In order to predict how the simulator would behave if we had run it in full, we must first determine how the simulator responds to changes in parameter values for which we have trained the model.

Perhaps the simplest way to estimate the relationship between the singular values and their corresponding parameters would be to use a least squares approach.  That is fitting a multivariate normal linear model to these vectors, assuming independent errors, or,

\begin{equation}
\mathbf{S} \sim \text{N}(\mathbf{\Theta}^\intercal\mathbf{B},\Sigma),\nonumber
\end{equation}

\noindent where $\mathbf{S}$ is a two-dimensional array with rows equal to each of the $\mathbf{s_p}$ and $\Sigma$ is a diagonal matrix.

Initial inversions using the full least squares estimate failed to provide constraints on the parameters of interest (Swallow \textit{et al.}, unpublished data), presumably as there was too much flexibility in the model when allowing all fifteen parameters to contribute to all of the singular values individually.  In order to reduce the degrees of freedom in the modelling process, we choose to apply a covariate selection approach to ensure only the most important parameters are selected for each of the singular values.  In particular we use a multivariate linear regression to relate the parameters to the singular values and select the principal covariates for each of the singular values using Akaike's information criterion (AIC).

AIC is chosen over the Bayesian information criterion (BIC) as it is consistent for prediction, and the main aim of this is to form the best prediction of $H$, without running the full simulation, rather than necessarily a statistically consistent model \citep[e.g.][]{yang05}.

A linear model is fitted to each singular value for each site, with AIC used to select the principal parameters explaining variability across parameter runs.  The process is carried out with forward stepwise selection using the \pkg{stepAIC} function in \proglang{R}. This begins with a simple intercept-only model and progressively adds parameters in one by one until adding parameters fails to reduce the AIC sufficiently.  The \pkg{stepAIC} function is applied to each row of the $\mathbf{D}$ matrix in turn,

\begin{equation}
d_{sp}=\alpha + g_{\theta}(\mathbf{\theta}^{-})^\intercal\mathbf{\beta} + \epsilon, \nonumber
\end{equation}

\noindent where $\mathbf{\theta}^{-}$ is a (possibly empty) subset of the original parameters and $g_{\theta}(.)$ are the parameter-specific transformations given in Table \ref{tab:parranges}.  The site-specific parameters are only allowed to act on singular values corresponding to that site.  Therefore for each site and singular value a binary vector of indices is returned for which parameters are selected (if any), as are the corresponding coefficients.  Table \ref{tab:AICbin} shows the binary indices for the first few rows of $\mathbf{D}$ and the site-specific totals of singular values for which each parameter is selected.  As the intercept (INT) is always included, the total for this parameter relates to the number of singular values associated with that site, or equivalently the number of observations at that site.  Dividing the parameter totals by this value gives a proportion of all singular values for which that parameter is considered an important predictor.

\begin{table}[!h]
\centering
\resizebox{\columnwidth}{!}{%
\begin{tabular}{c|c||c|c|c|c|c|c|c|c|c|c|c|c|c|c|c|c|c}
\toprule
$i$ & $s$ & INT & MBL & FTT & UMM & BLHS & BLVS & BLHU & BLVU & LHS & LVS & LHU & LVU & BLD & X$_s$ & Y$_s$ & Z$_s$  \\
\midrule
1 & 1 & 1 & 1 & 1 & 1 & 0 & 0 & 0 & 0 & 0 & 0 & 0 & 1 & 1 & 0 & 1 & 0 \\ 
  2 & 1 & 1 & 1 & 1 & 0 & 1 & 0 & 0 & 1 & 0 & 0 & 0 & 1 & 1 & 0 & 1 & 0 \\ 
  3 & 1 & 1 & 1 & 1 & 0 & 0 & 1 & 0 & 1 & 0 & 0 & 0 & 0 & 1 & 0 & 0 & 1 \\ 
  4 & 1 & 1 & 1 & 1 & 1 & 0 & 1 & 0 & 0 & 0 & 0 & 0 & 0 & 1 & 0 & 0 & 0 \\ 
  5 & 1 & 1 & 1 & 1 & 1 & 0 & 0 & 0 & 0 & 0 & 0 & 0 & 1 & 1 & 0 & 1 & 0 \\ 
  \vdots & \vdots & \vdots & \vdots & \vdots & \vdots & \vdots & \vdots & \vdots & \vdots & \vdots & \vdots & \vdots & \vdots & \vdots & \vdots & \vdots & \vdots \\ 
\bottomrule
Total & 1 & 120 & 35 & 90 & 57 & 21 & 36 & 27 & 31 & 10 & 18 & 23 & 31 & 71 & 19 & 43 & 21 \\
 & 2 & 114 & 73 & 99 & 93 & 15 & 40 & 18 & 43 & 7 & 47 & 21 & 72 & 87 & 71 & 47 & 62 \\
  & 3 & 119 & 15 & 109 & 87 & 27 & 56 & 22 & 55 & 11 & 51 & 56 & 40 & 74 & 19 & 22 & 87 \\
   & 4 & 67 & 33 & 66 & 33 & 12 & 27 & 9 & 9 & 17 & 6 & 13 & 14 & 31 & 23 & 28 & 51 \\
\bottomrule
\end{tabular} }
\caption{Example stepwise selection indices and site-specific totals. The intercept (INT) is always present, so the totals under this column represent the total number of singular values (and hence observations) for that site.}
\label{tab:AICbin}
\end{table}

\begin{figure}[!ht]
\centering
\includegraphics[height=0.4\textheight]{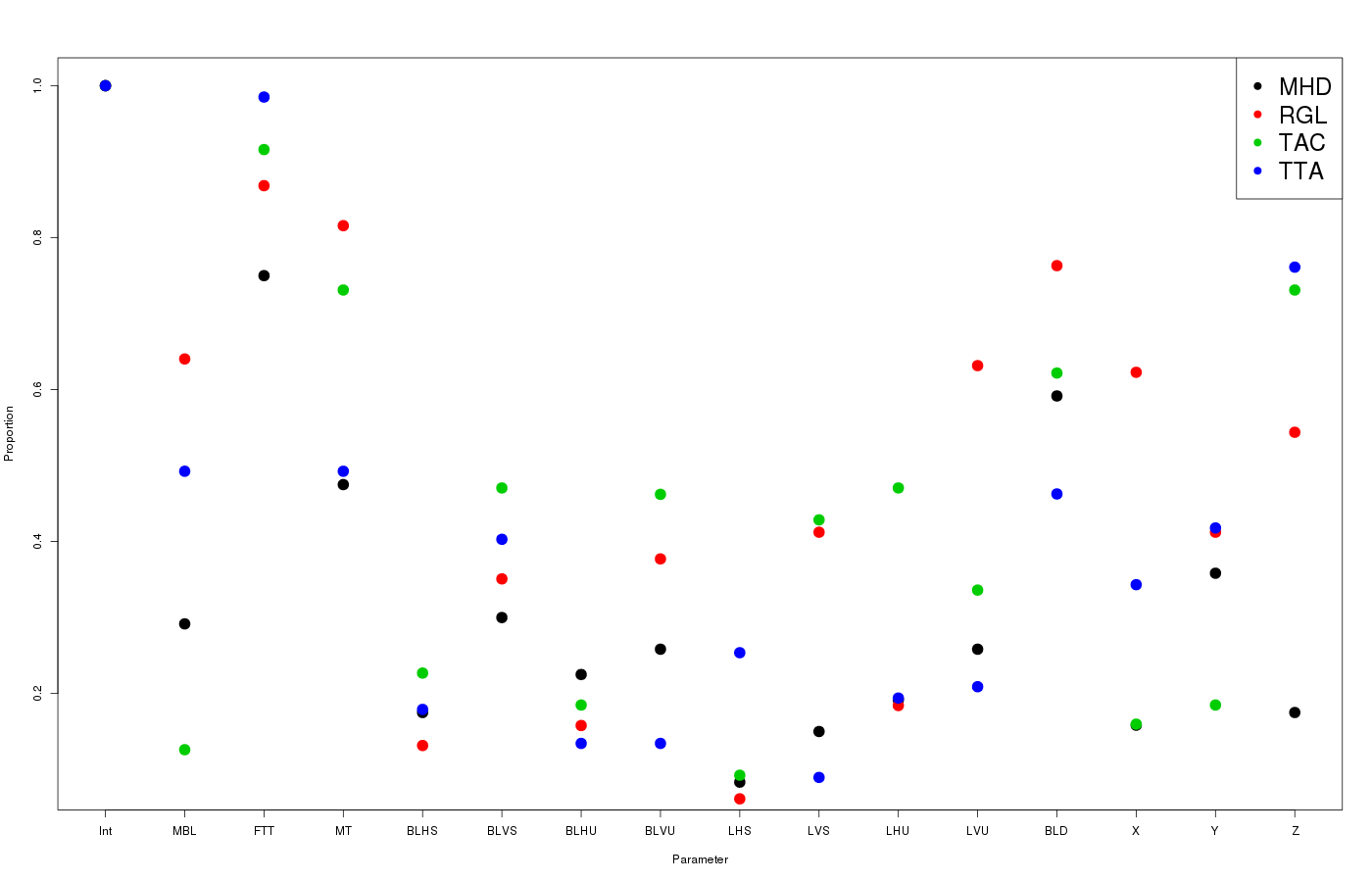}
\caption{Proportion of singular values for each site (Mace Head: black; Ridge Hill: red; Tacolneston: green; Angus: blue) where each parameter is selected through forward stepwise selection by AIC.}
\label{fig:propabs}
\end{figure}

\begin{figure}[!ht]
\centering
\includegraphics[height=0.4\textheight]{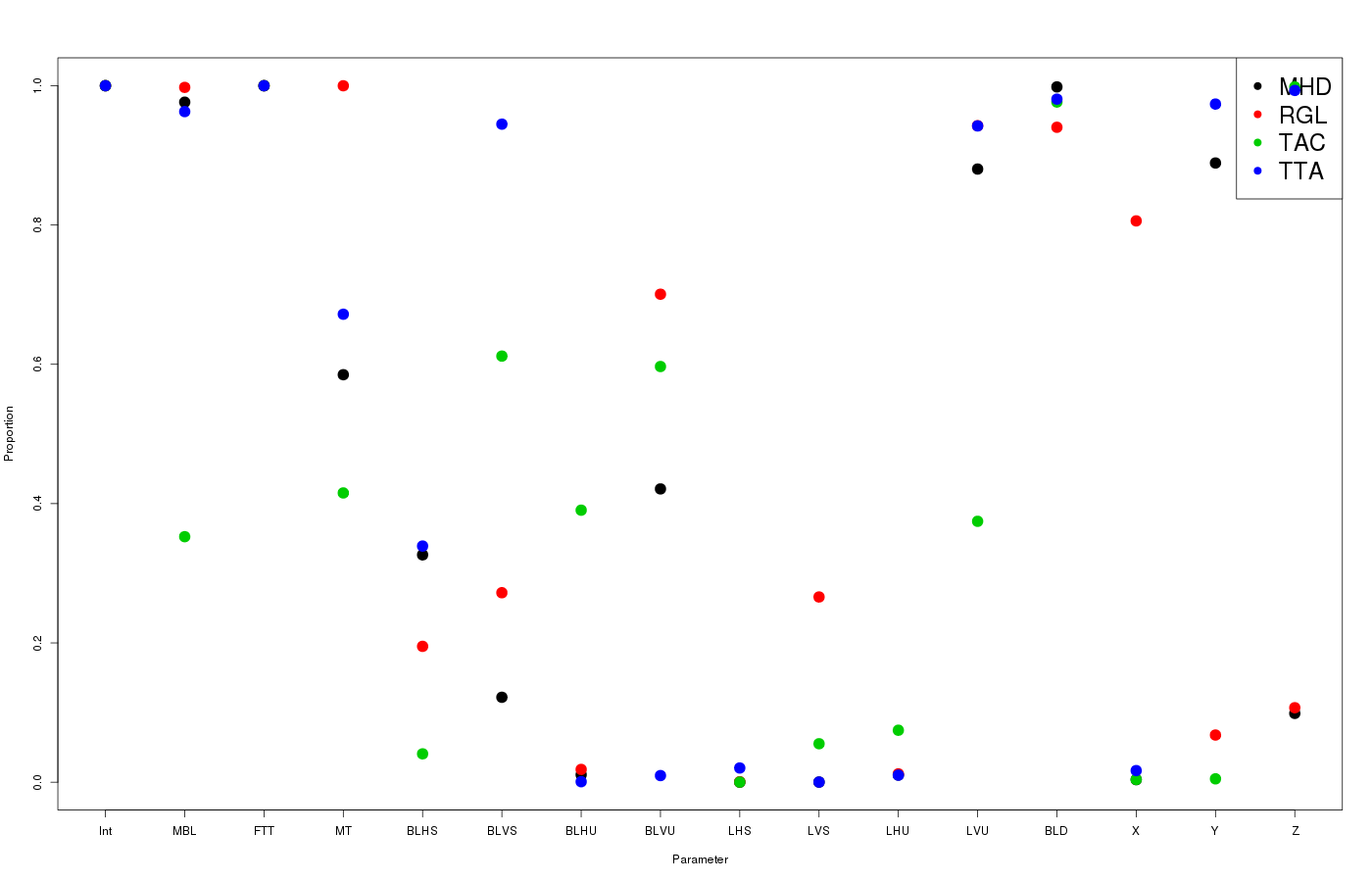}
\caption{Proportion of singular values for each site (Mace Head: black; Ridge Hill: red; Tacolneston: green; Angus: blue) where each parameter is selected through forward stepwise selection by AIC, weighted by the sample-average of $d_{si}$.}
\label{fig:propscaled}
\end{figure}

The proportion of singular values for which each parameter is selected by AIC can indicate to what extent each parameter is important in accounting for variation in the $H$ matrix.  Figure \ref{fig:propabs} shows the proportion of singular values segregated by site that each parameter is selected for.  Whilst this can give a good idea of how many of the singular values each parameter is considered important, it does not take into consideration the fact that each singular value accounts for a different proportion of the variability in $H^c$.  It is entirely feasible that a parameter that appears consistently for the 80\% of the lower singular values, for example, whilst these singular values may only account for 5\% of the overall variability.  Conversely, a parameter could be only selected for the first singular value, which may account for 80\% of the variance in $H^c$.  Figure \ref{fig:propscaled}, therefore, tries to avoid this discrepancy by normalising the contribution of each singular value to the proportions by the corresponding sample-averaged proportion of variation explained by that singular value.

\section{Estimation of parameters from observations}\label{sec:inversion}

In order to estimate the values of the parameters most consistent with the observational data, we incorporate the statistical model developed in Sections \ref{sec:training} to \ref{sec:statmod} into the Bayesian hierarchical model detailed in \citet{ganesan14} and estimate the values of the parameters using a MCMC framework. The MCMC framework uses using the Metropolis-Hastings (MH) algorithm to generate a Monte Carlo sample of parameter values from the posterior distributions of interest.  The algorithm was previously implemented for a similar study by \citet{ganesan14} and \citet{lunt16} but needs to be extended to incorporate uncertainty in the additional parameters $\mathbf{\theta}$.  

Prior distributions are specified on each of the parameters of interest, in particular uniform priors are specified on the unknown NAME parameters, with ranges equal to the plausible ranges from Table \ref{tab:parranges}.

The Metropolis-Hasting algorithm requires the specification of proposal variances to define how new parameter values are generated based on the value at the current iteration. Within this study, these variances are tuned using an adaptive batch pilot tuning algorithm, in which the acceptance rate for each parameter is monitored for a fixed number of iterations.  If the proportion of iterations within that batch period where a given parameter is accepted is greater than a pre-specified value, the proposal for that parameter is increased in magnitude.  If the acceptance rate is below another pre-specified value, then the proposal is reduced.  This proved a satisfactory approach to providing good mixing within the MCMC chains, confirmed through post-analysis visual checks of parameter traceplots.

\subsection{Reconstructing $H$}
At each iteration of the Markov chain, a single update Metropolis-Hastings algorithm is used to update the parameters of $\theta$ in turn.  A new value is proposed by adding a random draw from a normal distribution centred on the current value with a tuned proposal standard deviation as described above.  This proposed value of the parameter is transformed by the relevant transformation, and the new $\theta$ vector is multiplied by $\mathbf{B}, \mathbf{U}$ and $ \mathbf{V}^{\intercal}$ in turn to generate a pseudo-$H^c$ corresponding to this new value.  The mean matrix $H_m$ is then added to give a full estimate of $H$.

\begin{equation}
H_{[s]}=\mathbf{U_{0,s}} \{diag(\mathbf{B}g_{\theta}(\mathbf{\theta})) \}\mathbf{V_{0,s}}^{\intercal}+H_{m,[s]}   \nonumber
\end{equation}

\noindent This new representation of $H$ is incorporated into Equation \ref{eqn:inversion}, and we now estimate simultaneously both the fluxed $\mathbf{x}$ and the NAME parameters $\theta$.

The acceptance probability is calculated and the parameter value is accepted with this probability as per the Metropolis-Hastings algorithm.

\section{Results}
The MH algorithm was run for 100,000 iterations with the first half discarded as burn-in.  The remaining sample was thinned by a factor of 10 giving a final posterior sample of 5,000.  Posterior summary statistics for each NAME input parameter are given in Table \ref{tab:postres} as well as the estimate of annual methane emissions for the UK under the new method and the inversion where $H$ is fixed.  There is a moderate increase in the estimate of the UK total methane emissions when uncertainty in $H$ is incorporated.  The biggest difference is seen in the increase in the upper bound of the 90\% credible interval, being almost 50\% higher than in the analysis without uncertainty in $H$.

\begin{table}[!ht]
\centering
\resizebox{\textwidth}{!}{\begin{tabular}{lcc}
\toprule 
Parameter/quantity & Posterior mean & 90\% credible interval \\
\midrule
MBL (m)  & 73.35 & (44.59,98.07) \\
FTT ($\sigma_u$/$\sigma_w$) & 0.88 & (0.74,0.99) \\
UMM & 0.57 & (0.28,0.81) \\
BLHS & 1.49 & (0.68,2.25) \\
BLVS & 1.14 & (0.63,1.64) \\
BLHU & 1.39 & (0.66,2.22) \\
BLVU & 1.13 & (0.63,1.64) \\
LHS & 0.64 & (0.17,1.06) \\
LHV & 0.57 & (0.14,1.04) \\
LHU & 0.62 & (0.16,1.06) \\
LVU & 0.64 & (0.17,1.06) \\
BLD (m) & 1.10 & (0.60,1.48) \\
Z$_\text{MHD}$ (m) & -0.001 & (-0.04,0.04) \\
Z$_\text{RGL}$ (m) & 0.001 & (-0.04,0.05) \\
Z$_\text{TAC}$ (m) & -0.001 & (-0.04,0.04) \\
Z$_\text{TTA}$ (m) & 0.002 & (-0.04,0.05) \\
X$_\text{MHD}$ & -0.001 & (-0.05,0.04) \\
X$_\text{RGL}$ ($\si{\degree}$) & 0.001 & (-0.04,0.05) \\
X$_\text{TAC}$ ($\si{\degree}$) & -0.002 & (-0.05,0.04) \\
X$_\text{TTA}$ ($\si{\degree}$) & 0.000 & (-0.04,0.05) \\
Y$_\text{MHD}$ ($\si{\degree}$) & 46.34 & (-5.37,101.29) \\
Y$_\text{RGL}$ ($\si{\degree}$) & 5.03 & (-85.04,143.32) \\
Y$_\text{TAC}$ ($\si{\degree}$) & 22.09 & (-89.93,165.75) \\
Y$_\text{TTA}$ ($\si{\degree}$) & 150.13 & (-126.04,315.48) \\
UK total emissions w/o uncertainty (Tg/yr) & 2.19 & (2.02,2.39) \\
UK total emissions with uncertainty (Tg/yr) & 2.33 & (2.14,3.31) \\
\bottomrule
\end{tabular}}
\caption{Posterior summary statistics for the hierarchical Bayesian inversion.}
\label{tab:postres}
\end{table}

\begin{figure}[!ht]
\centering
\resizebox{!}{0.15\textheight}{
\includegraphics[width=0.4\textwidth]{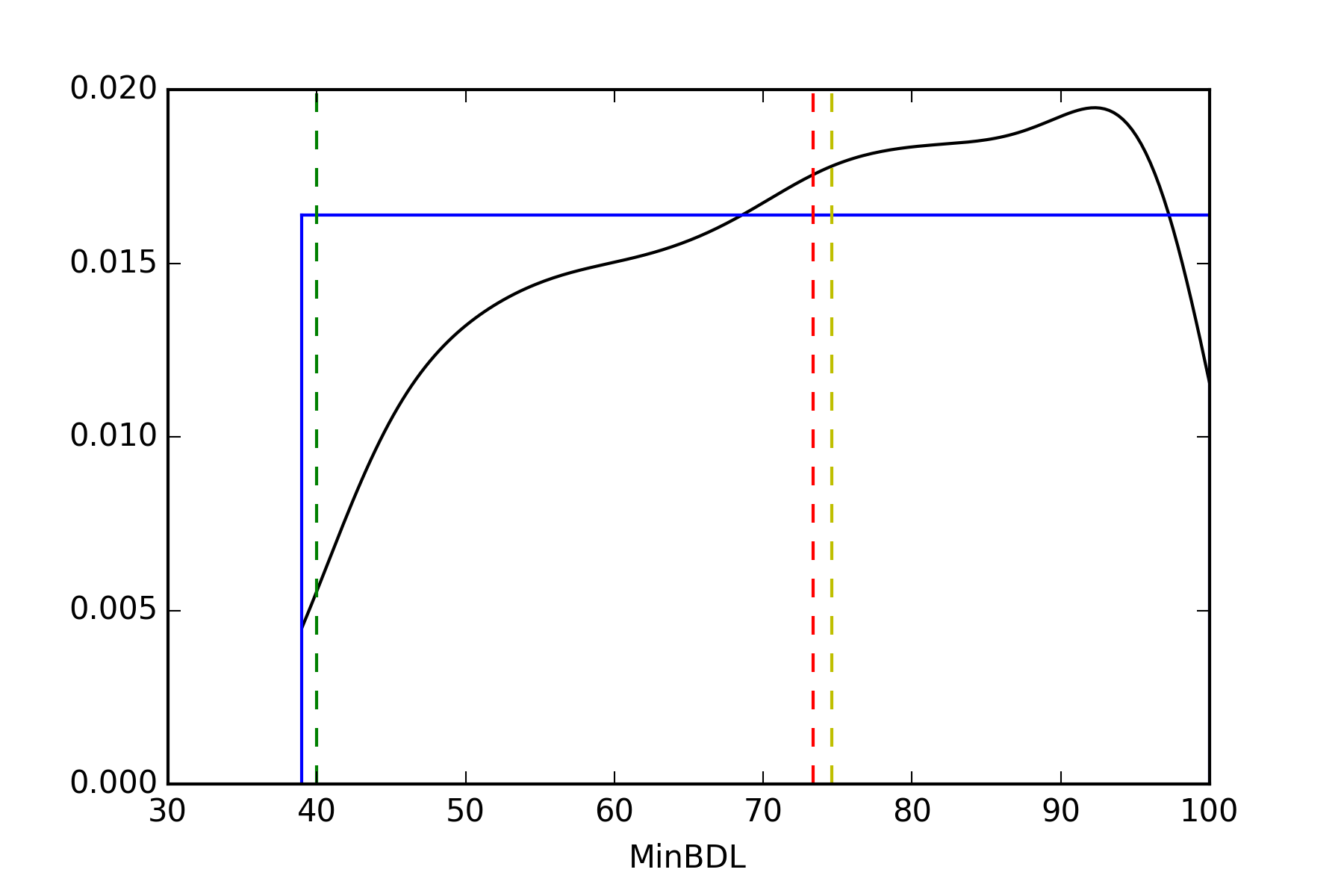}
\includegraphics[width=0.4\textwidth]{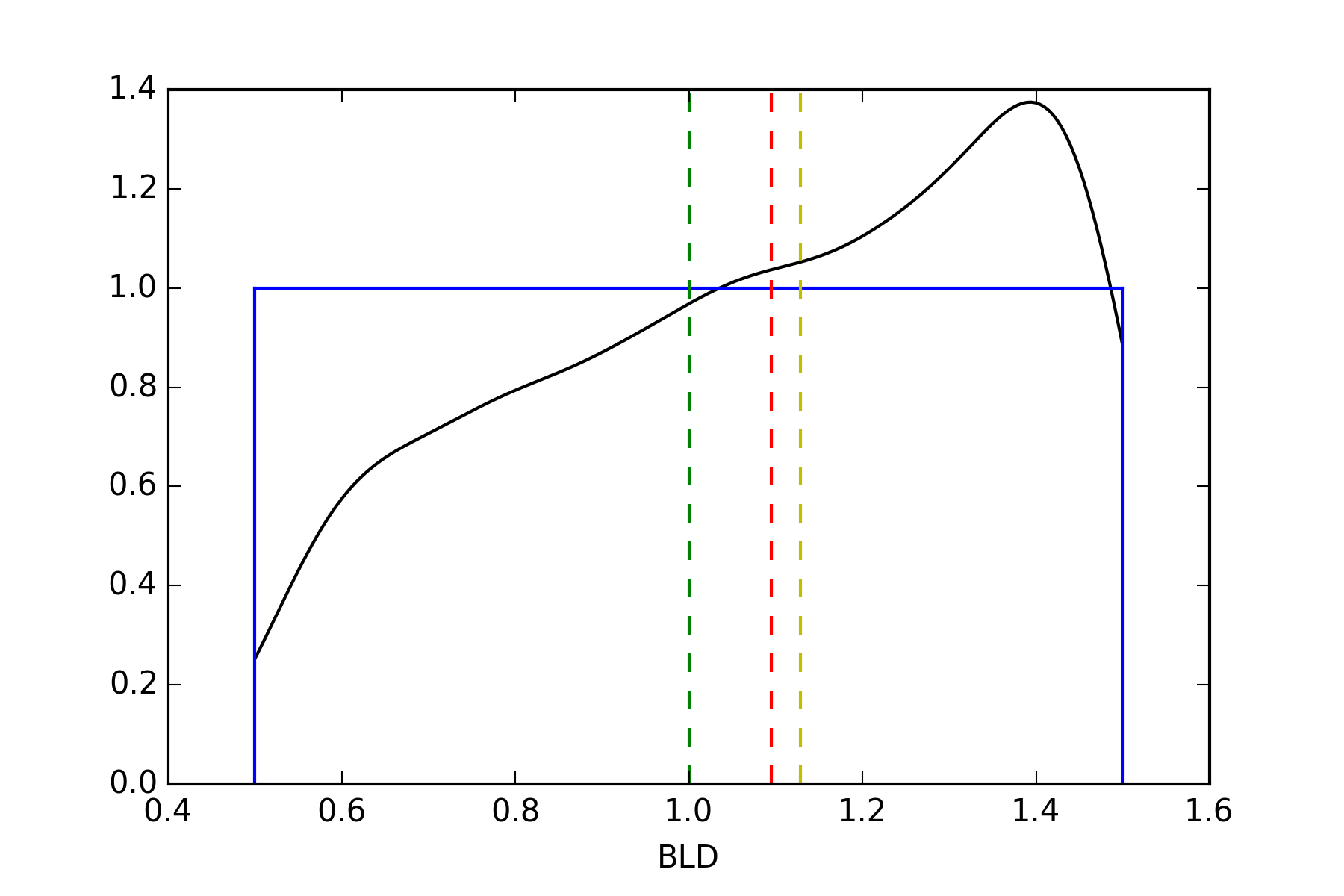}}\\
\resizebox{!}{0.15\textheight}{
\includegraphics[width=0.4\textwidth]{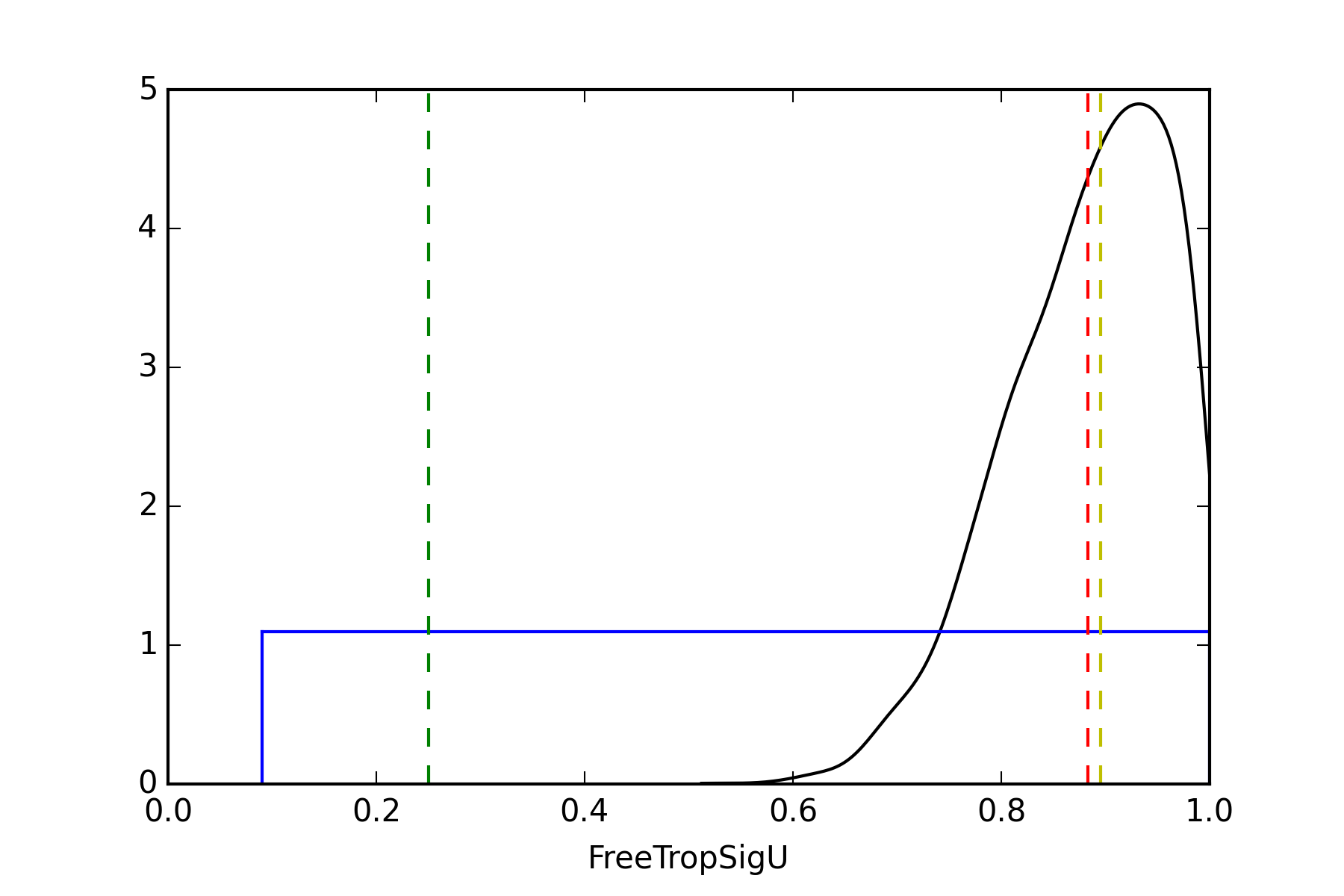}
\includegraphics[width=0.4\textwidth]{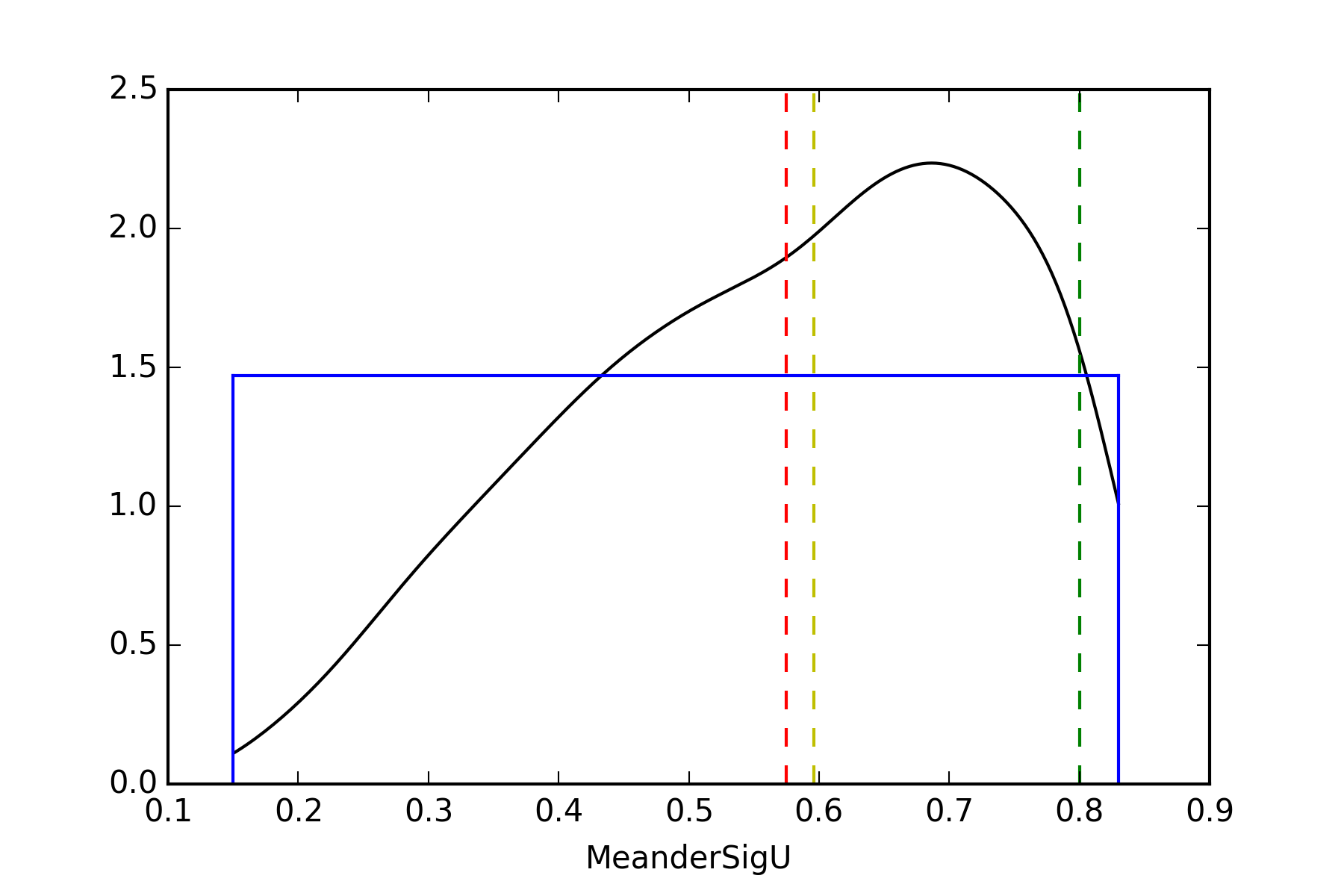}}\\
\resizebox{!}{0.15\textheight}{
\includegraphics[width=0.4\textwidth]{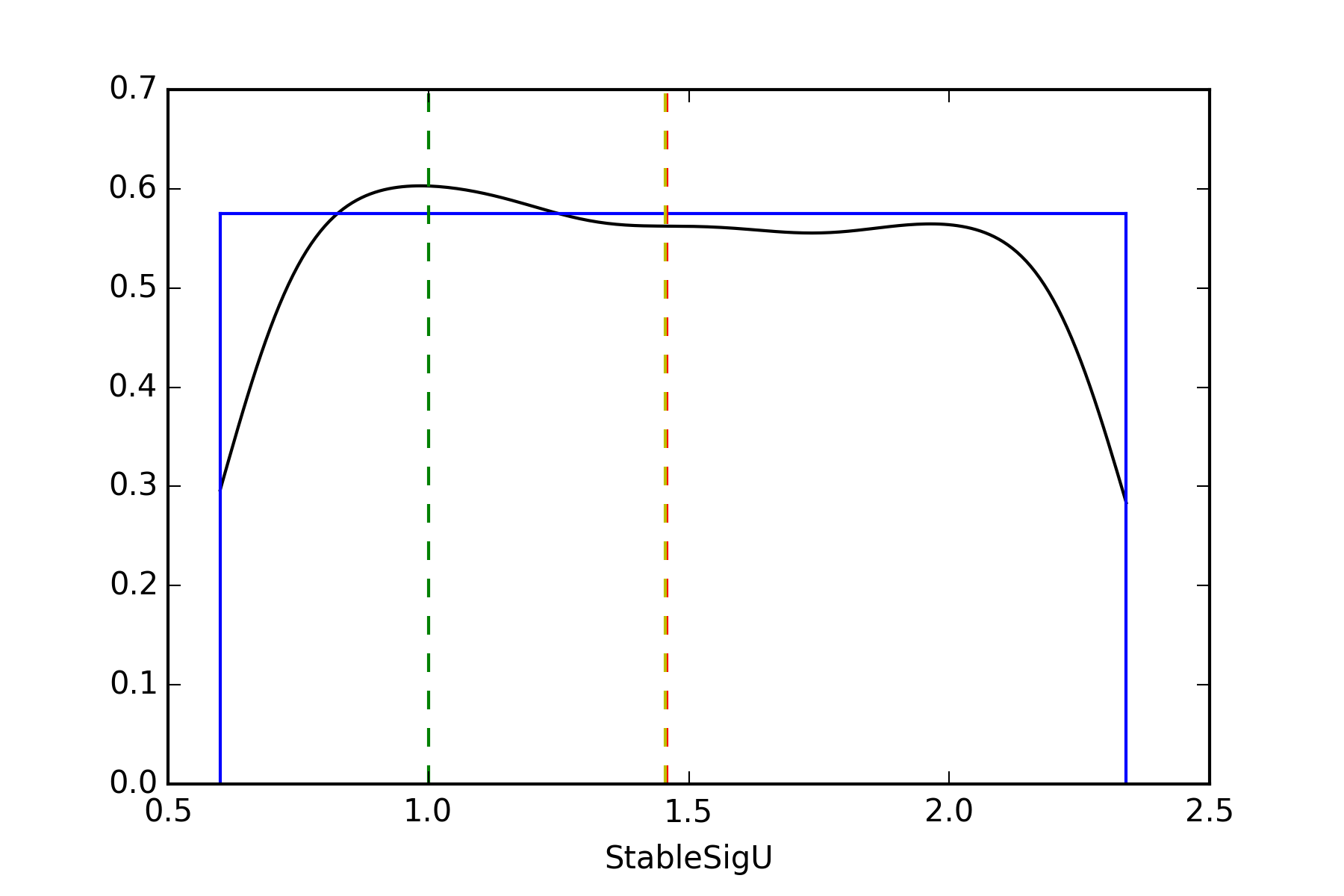}
\includegraphics[width=0.4\textwidth]{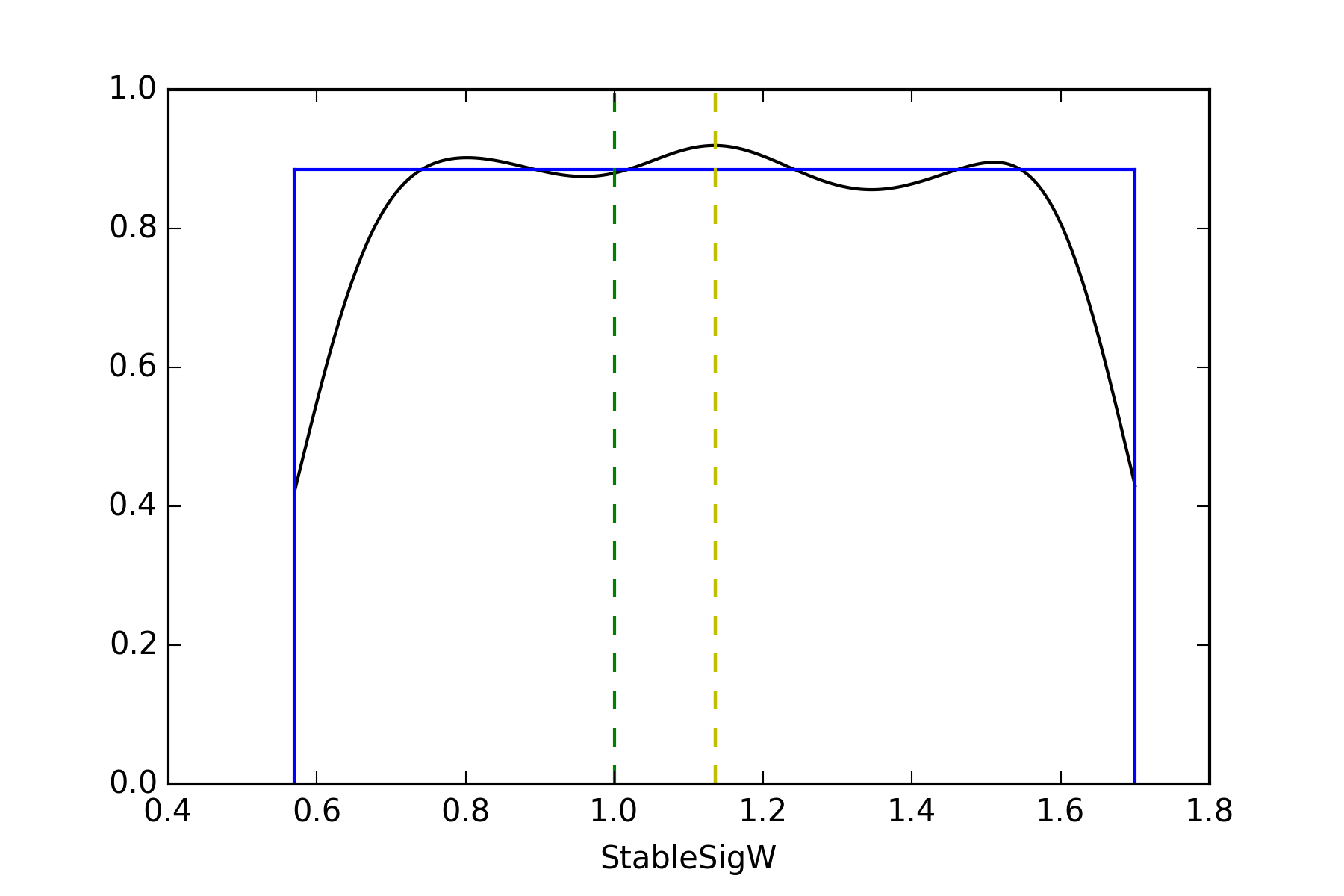}}\\
\resizebox{!}{0.15\textheight}{ 
\includegraphics[width=0.4\textwidth]{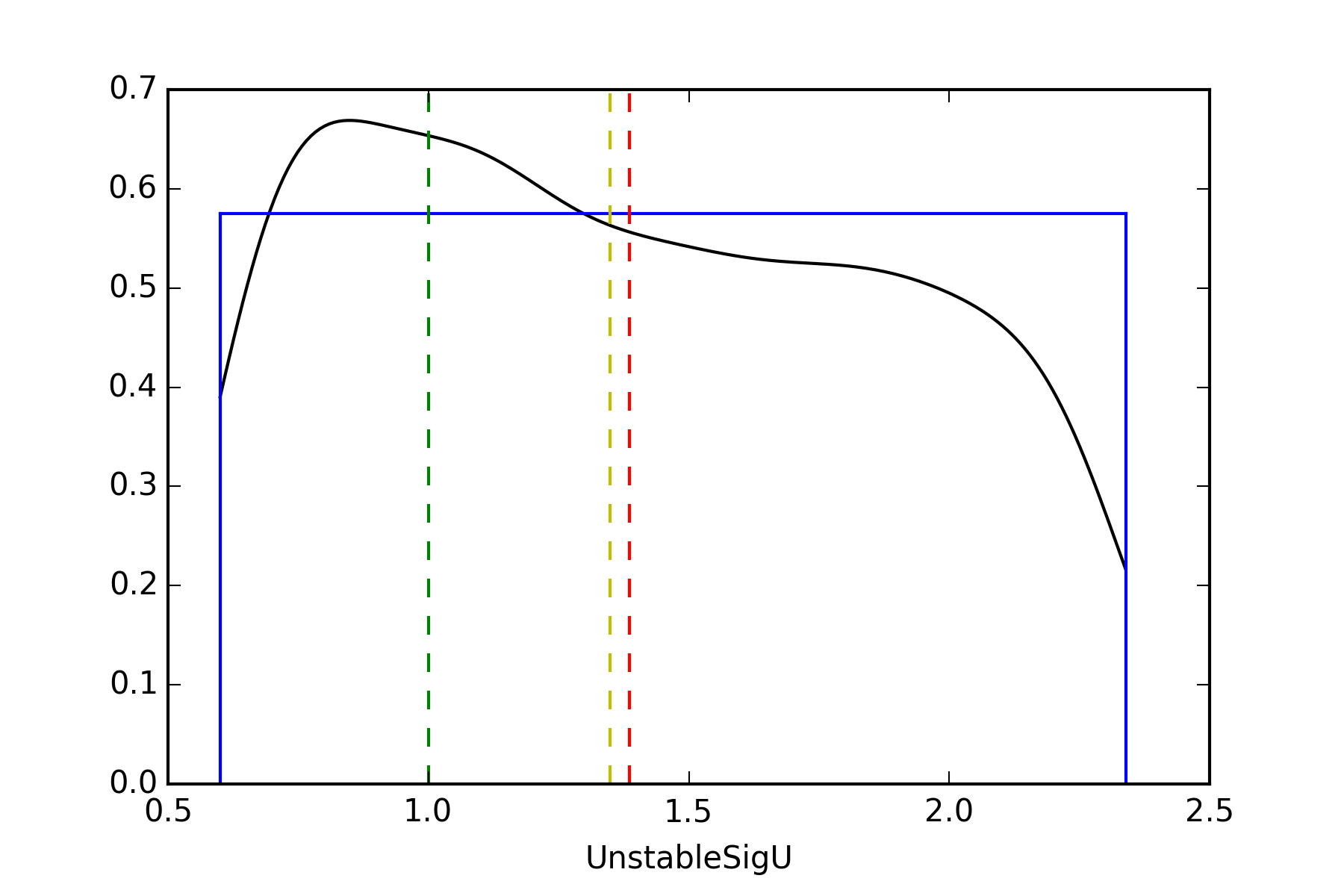}
\includegraphics[width=0.4\textwidth]{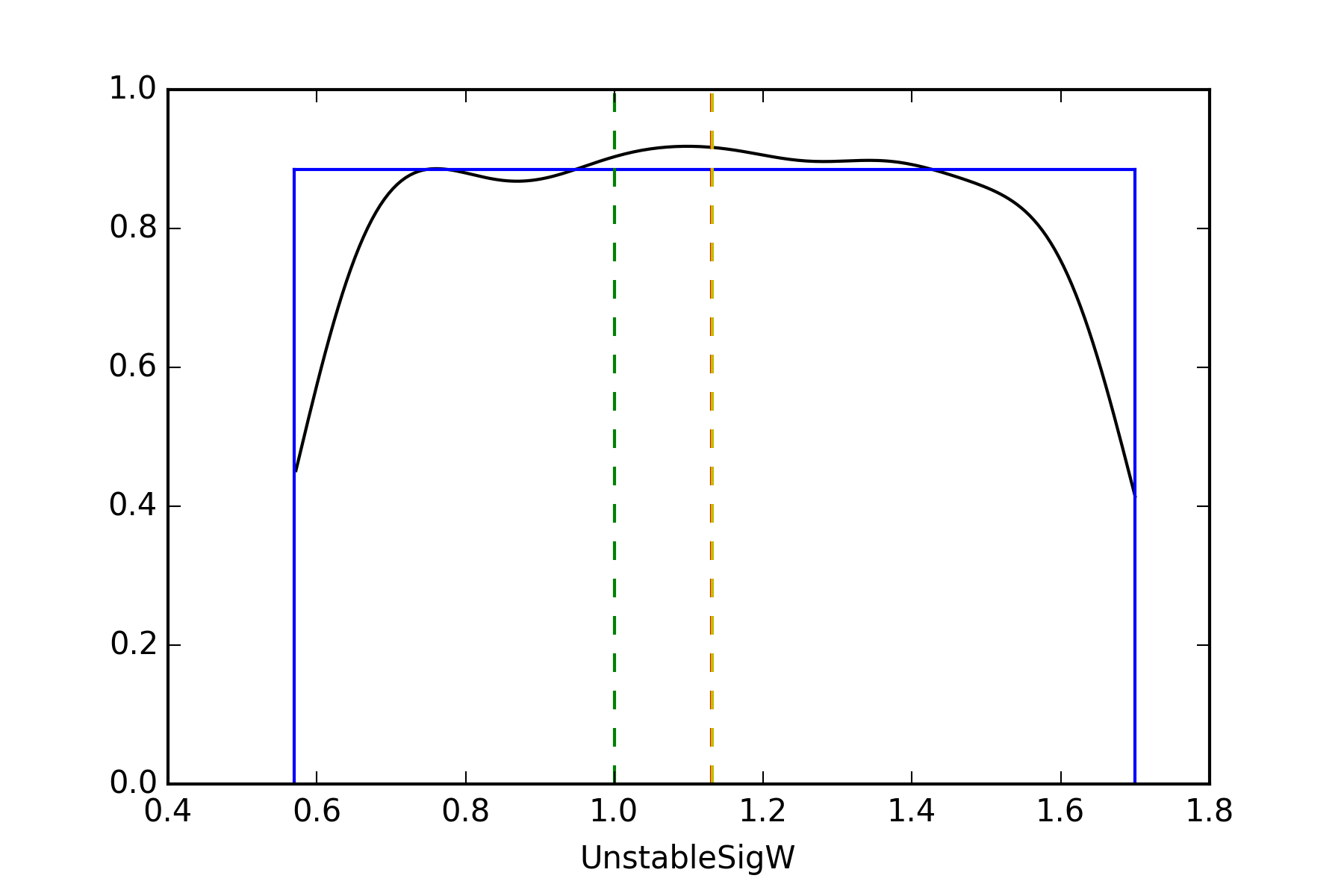}}\\
\resizebox{!}{0.15\textheight}{
\includegraphics[width=0.4\textwidth]{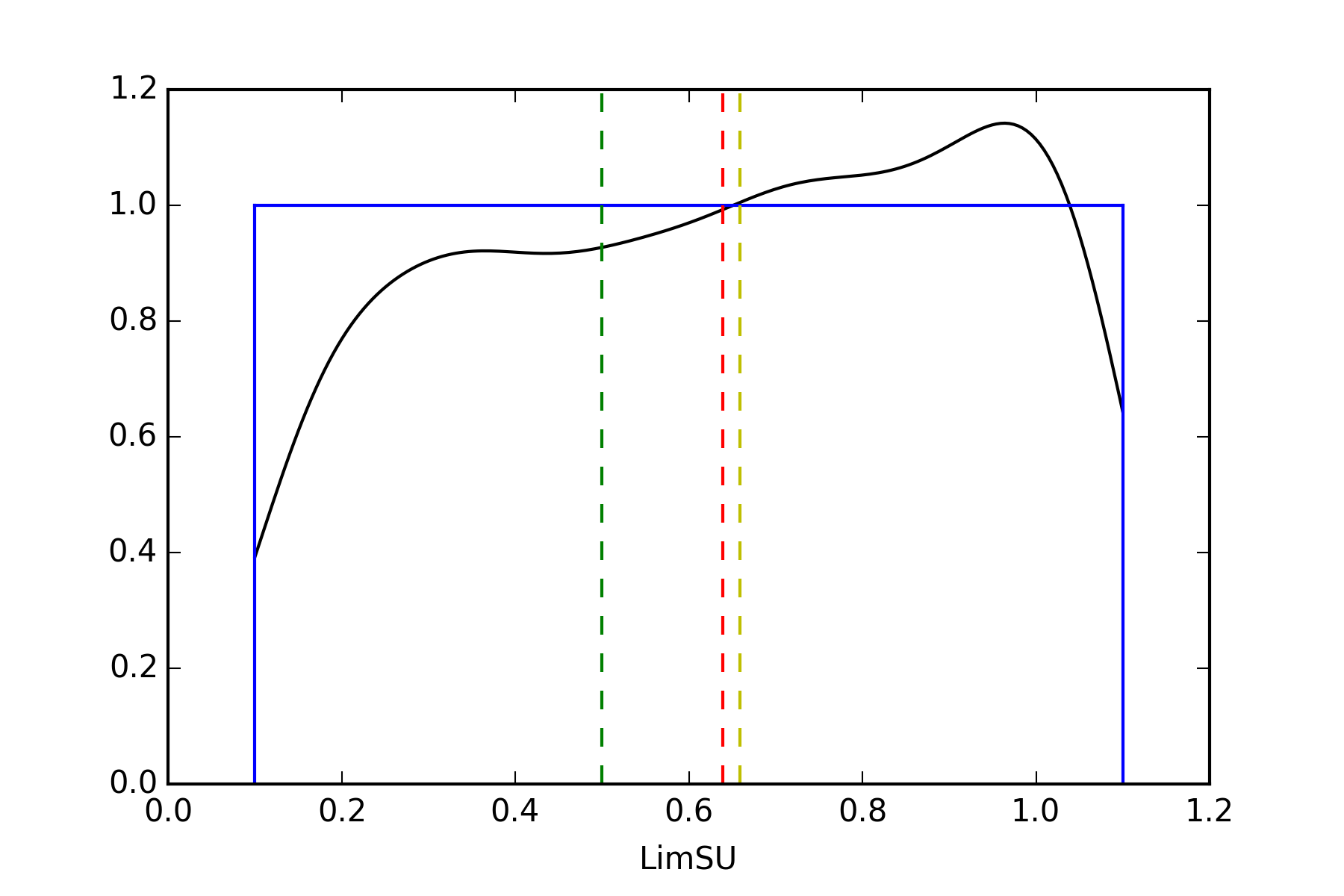}
\includegraphics[width=0.4\textwidth]{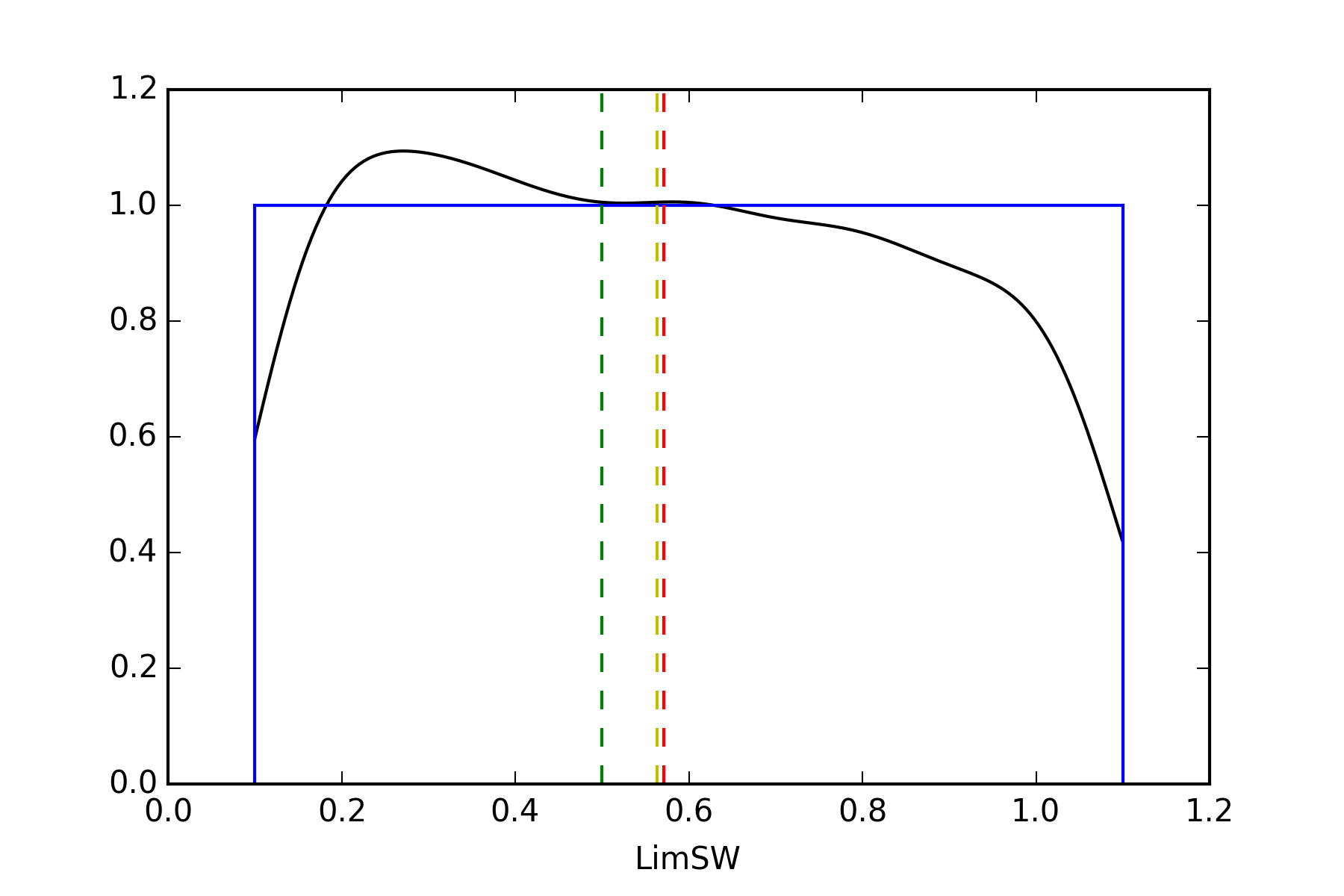}}\\
\resizebox{!}{0.15\textheight}{ 
\includegraphics[width=0.4\textwidth]{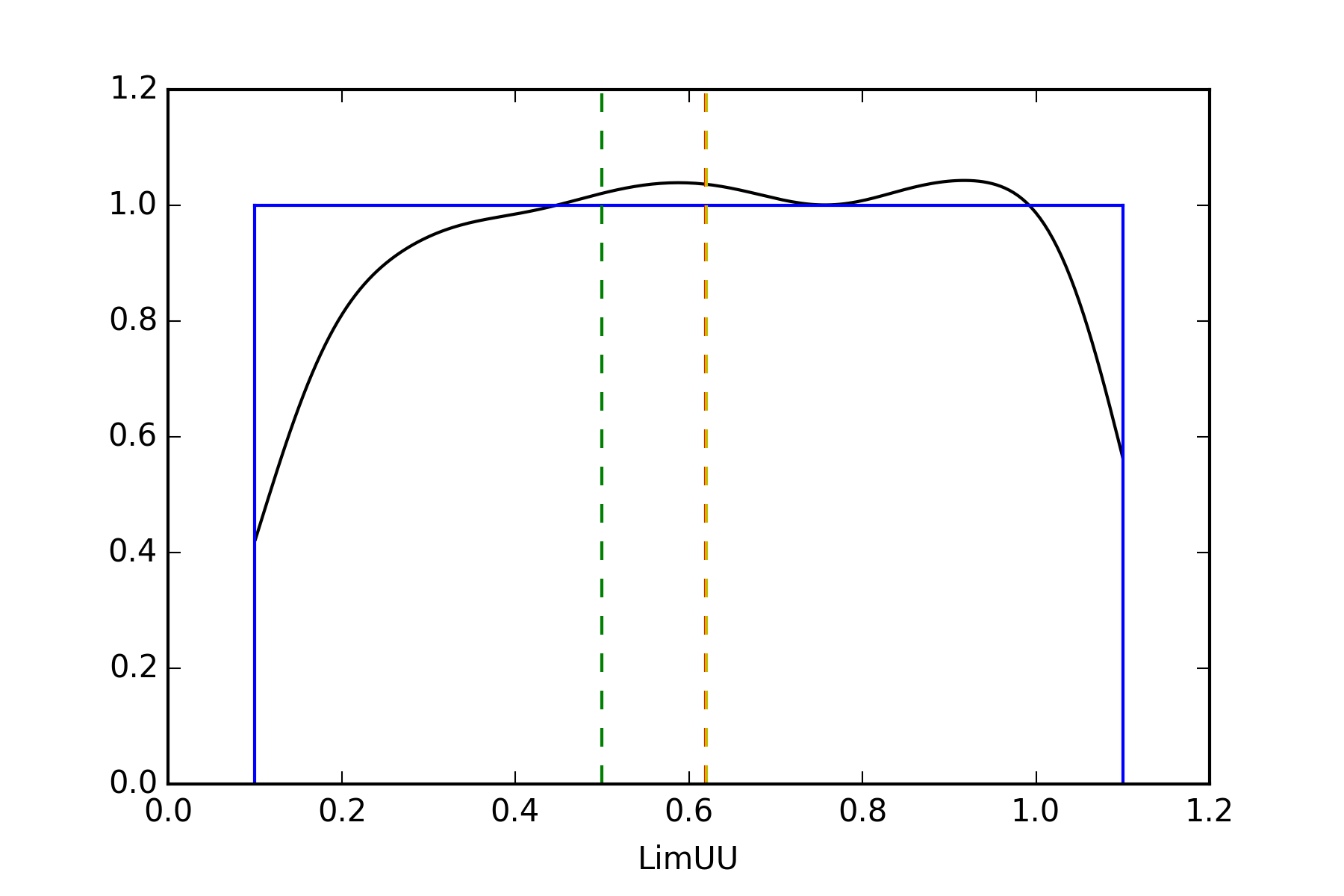}
\includegraphics[width=0.4\textwidth]{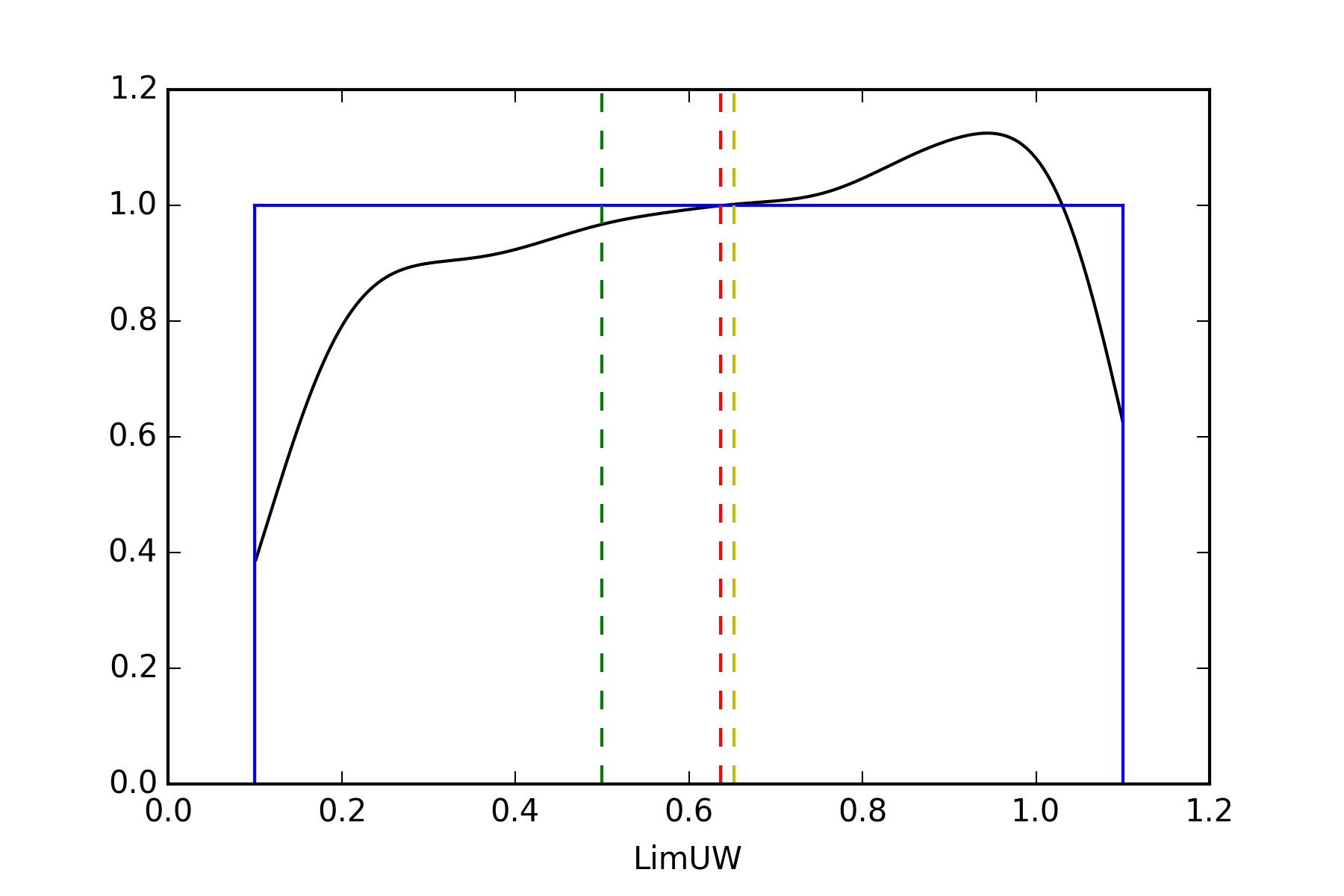}}
\caption{Posterior marginal distributions for the site-invariant parameters (black). Prior distributions are shown (blue), posterior mean (red) and median (yellow) and the default fixed parameter value (green).}
\label{fig:postmarinv}
\end{figure}

\begin{figure}[!ht]
\centering
\resizebox{!}{0.15\textheight}{
\includegraphics[width=0.4\textwidth]{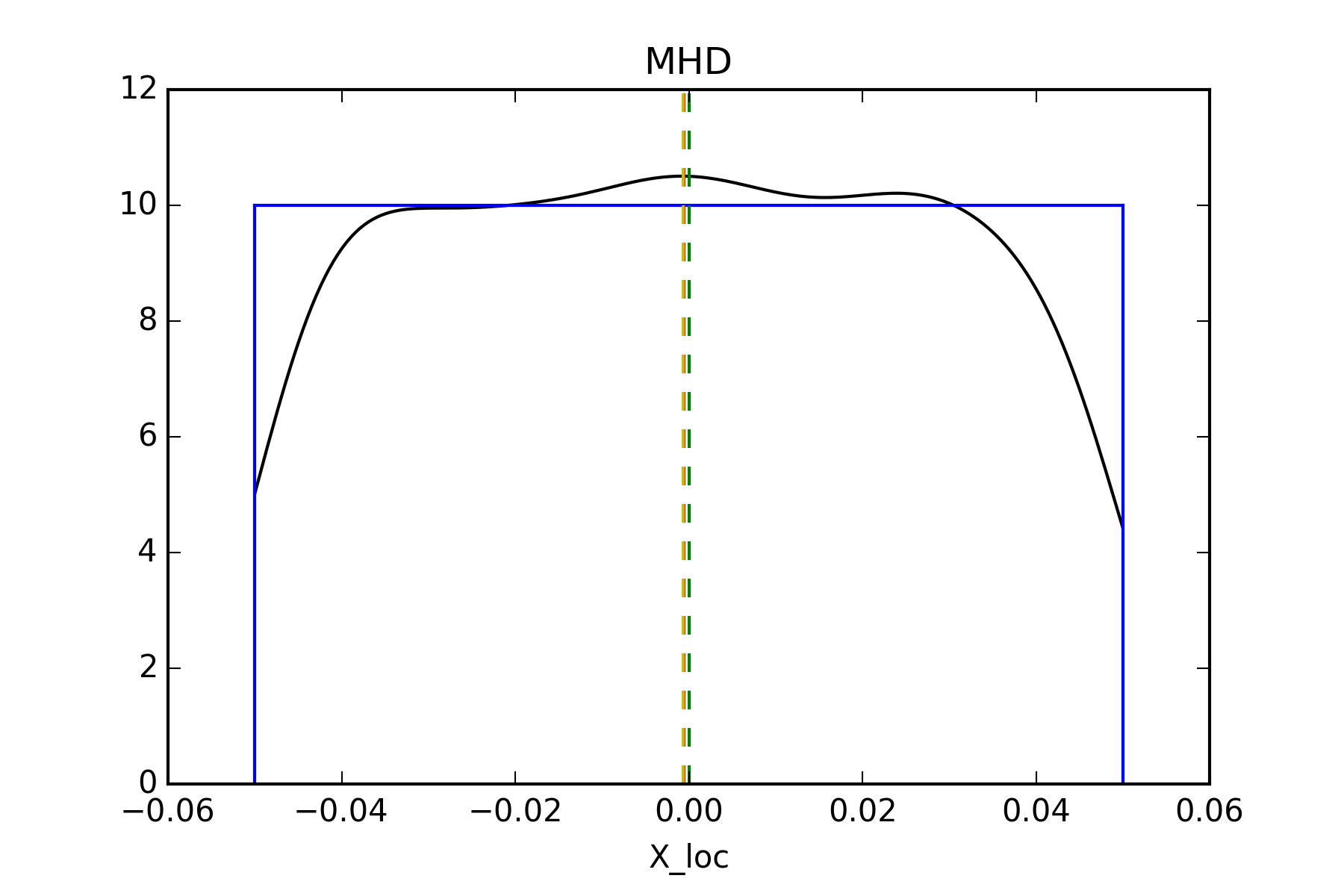}
\includegraphics[width=0.4\textwidth]{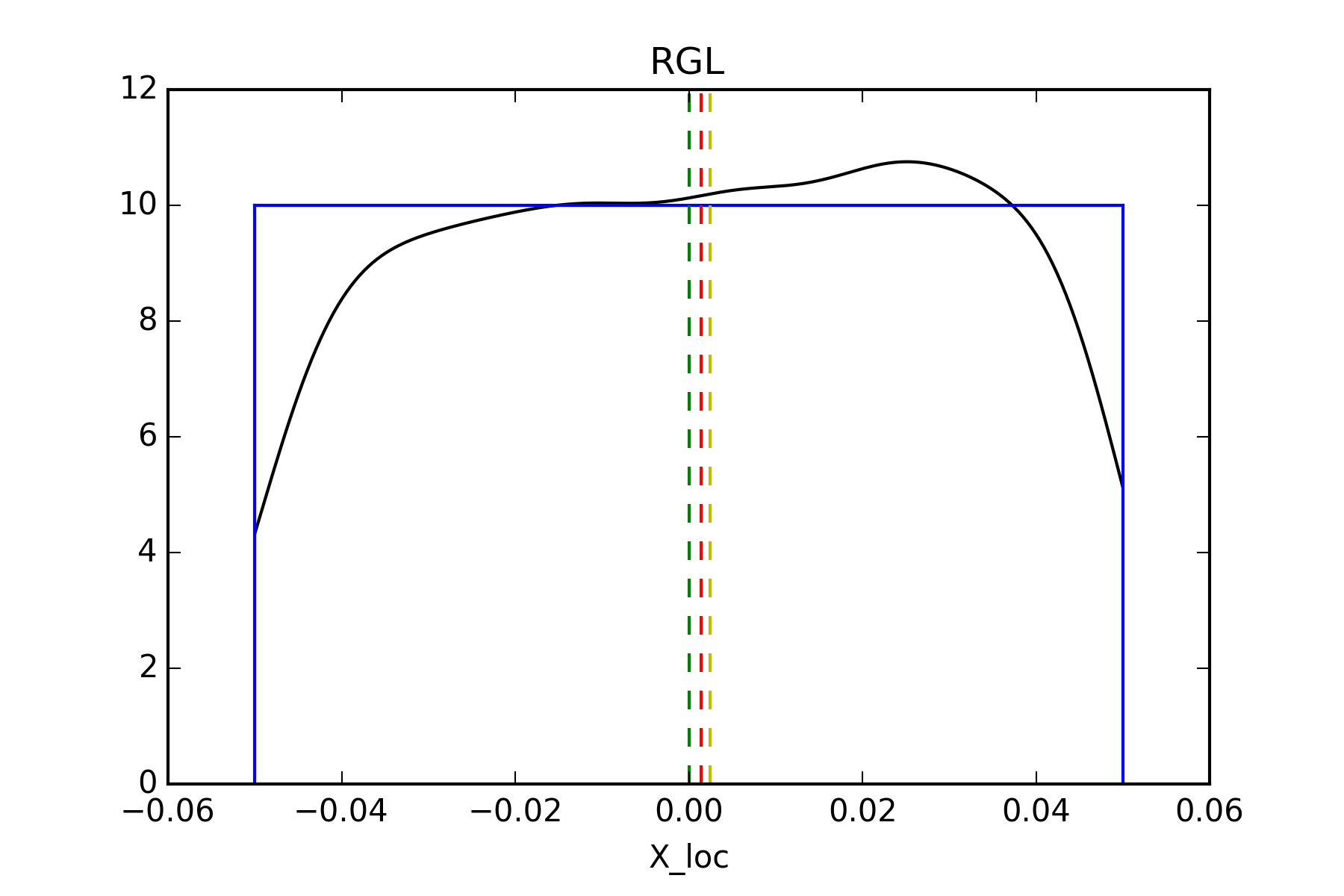}}\\
\resizebox{!}{0.15\textheight}{
\includegraphics[width=0.4\textwidth]{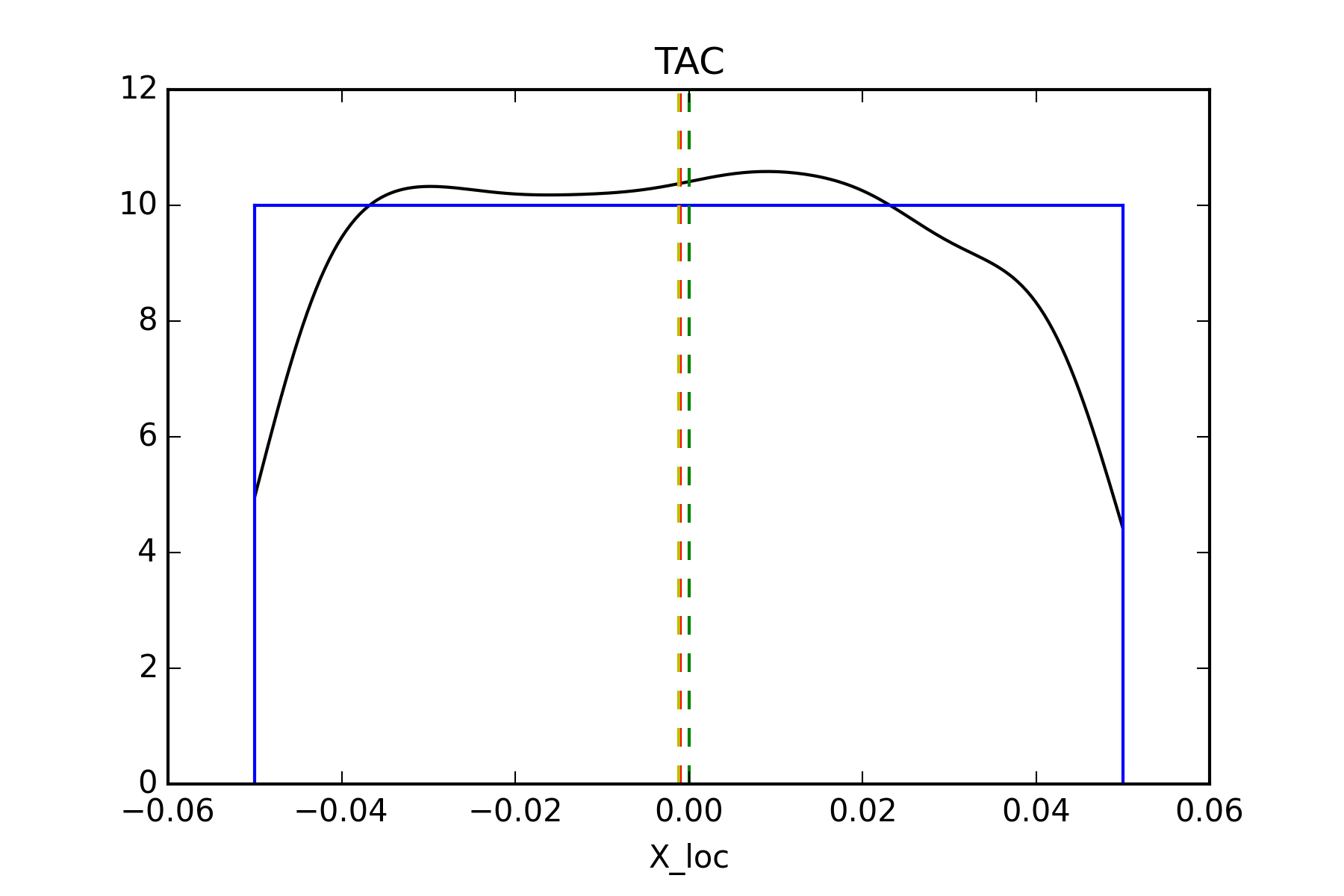}
\includegraphics[width=0.4\textwidth]{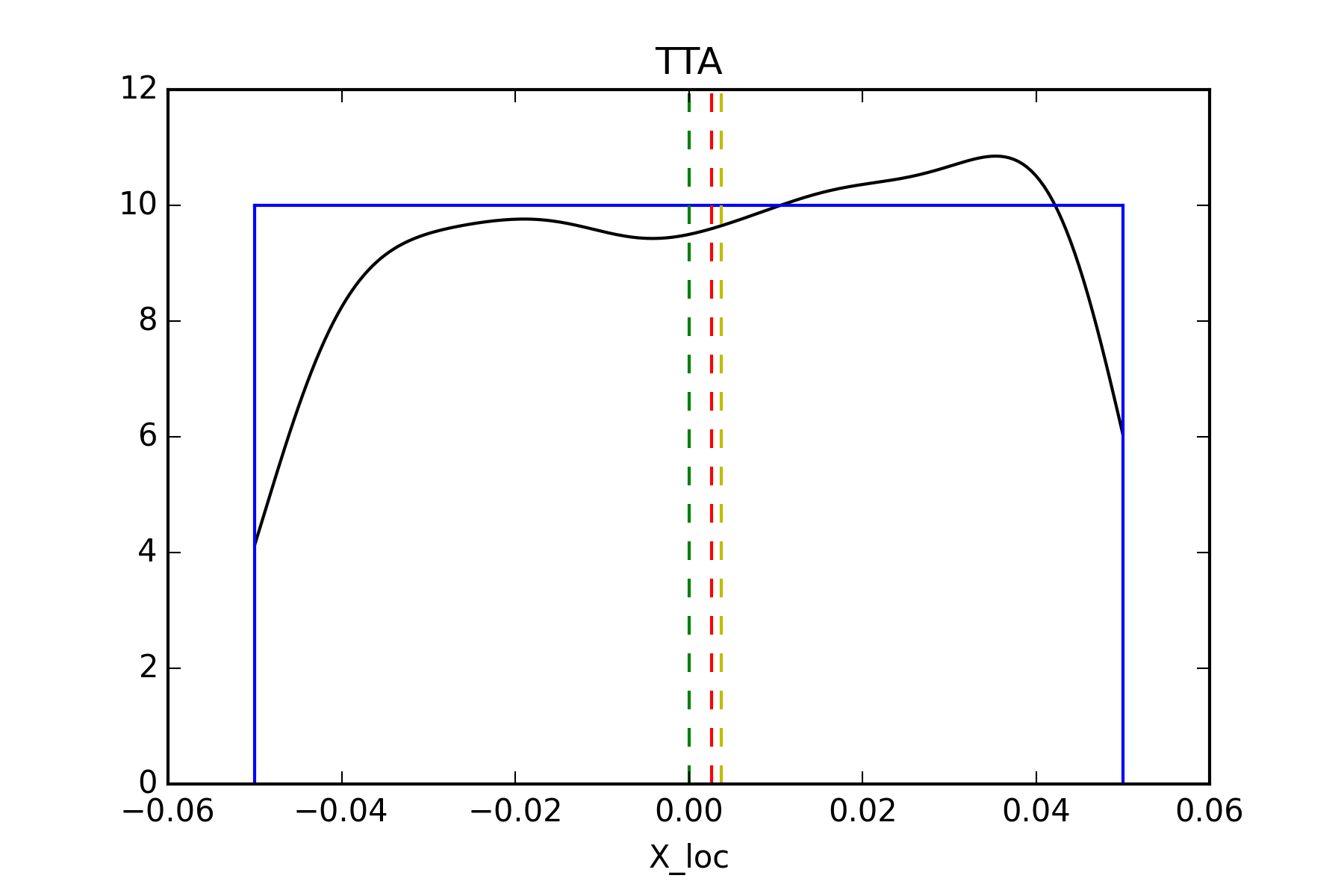}}\\
\resizebox{!}{0.15\textheight}{
\includegraphics[width=0.4\textwidth]{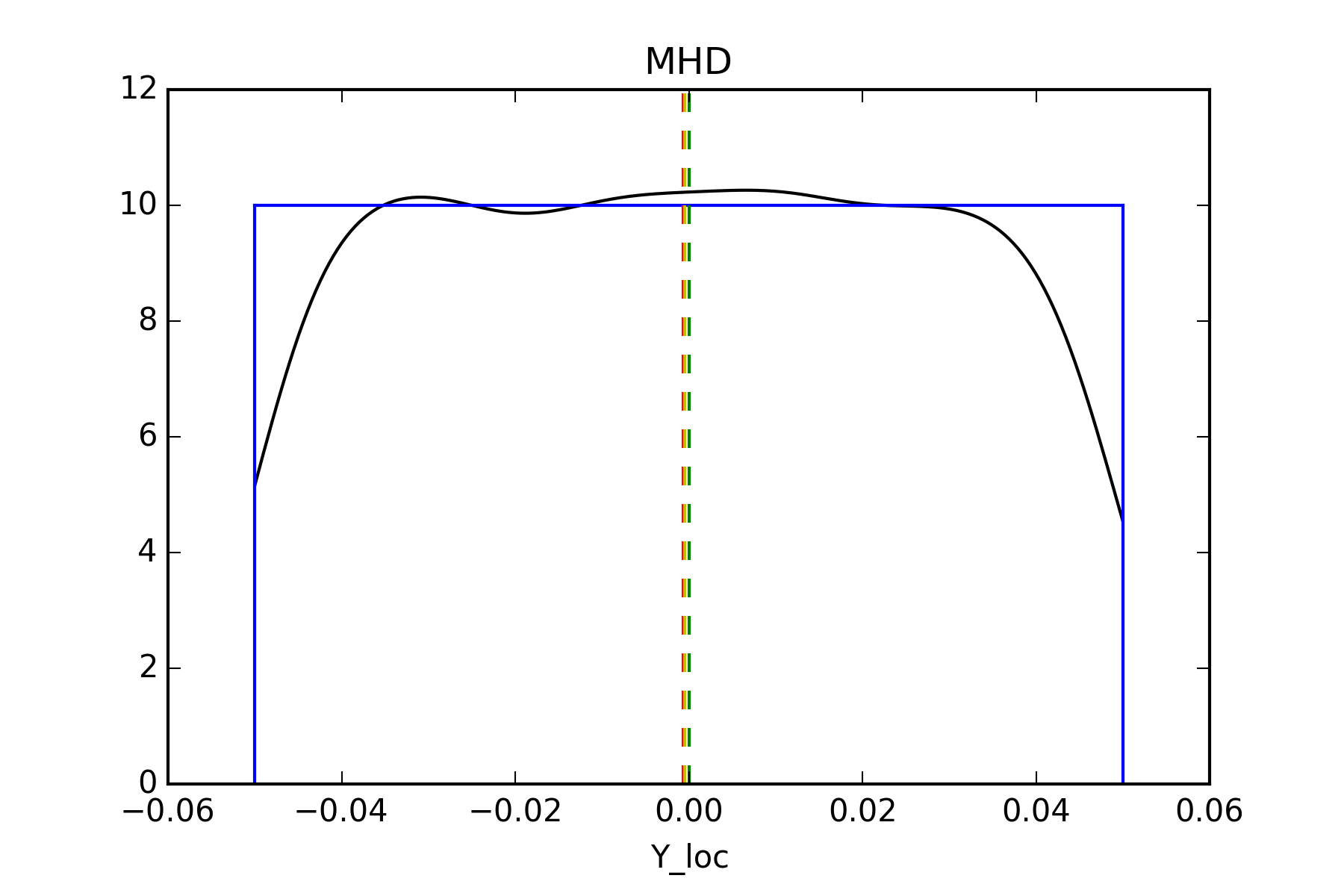}
\includegraphics[width=0.4\textwidth]{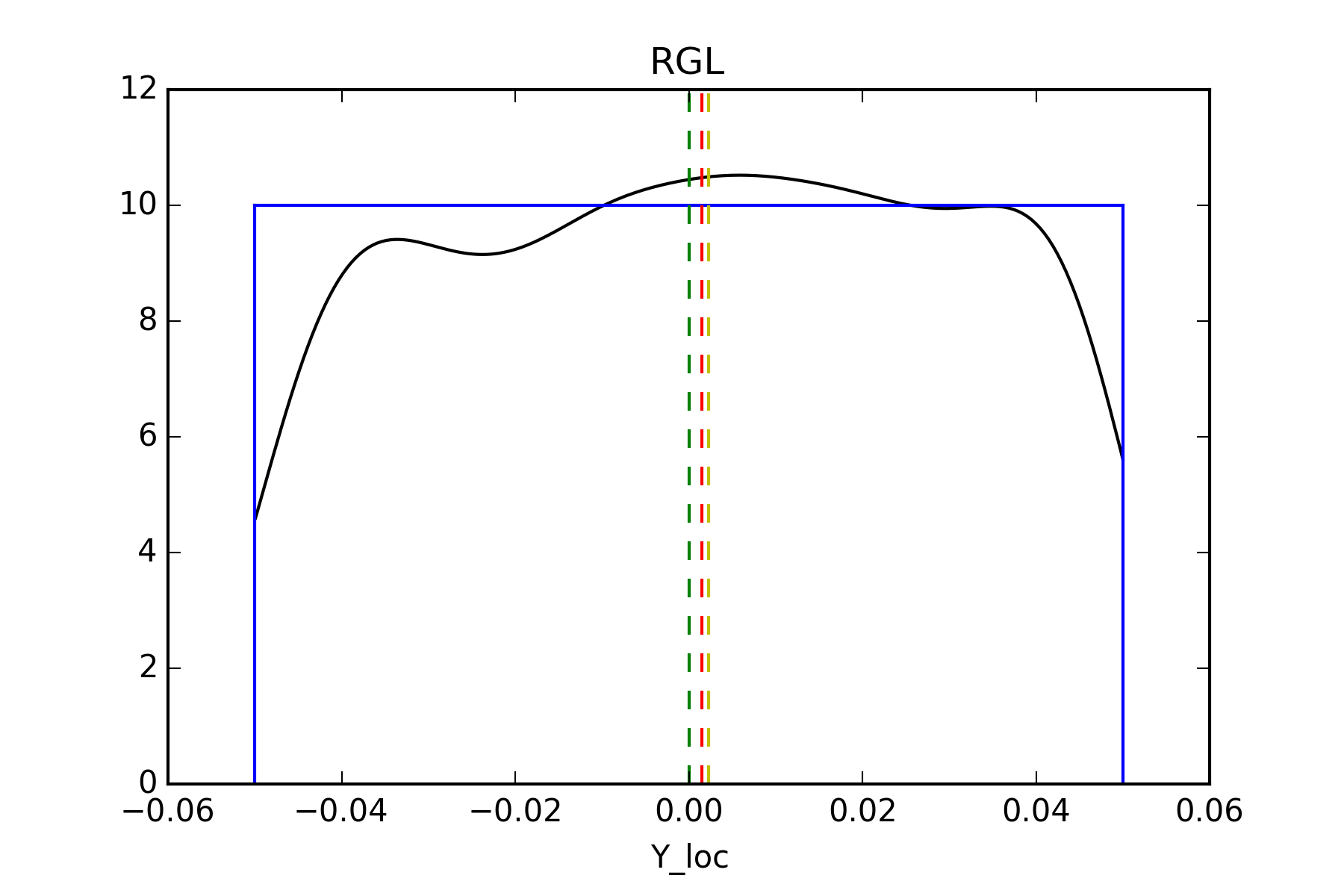}}\\
\resizebox{!}{0.15\textheight}{
\includegraphics[width=0.4\textwidth]{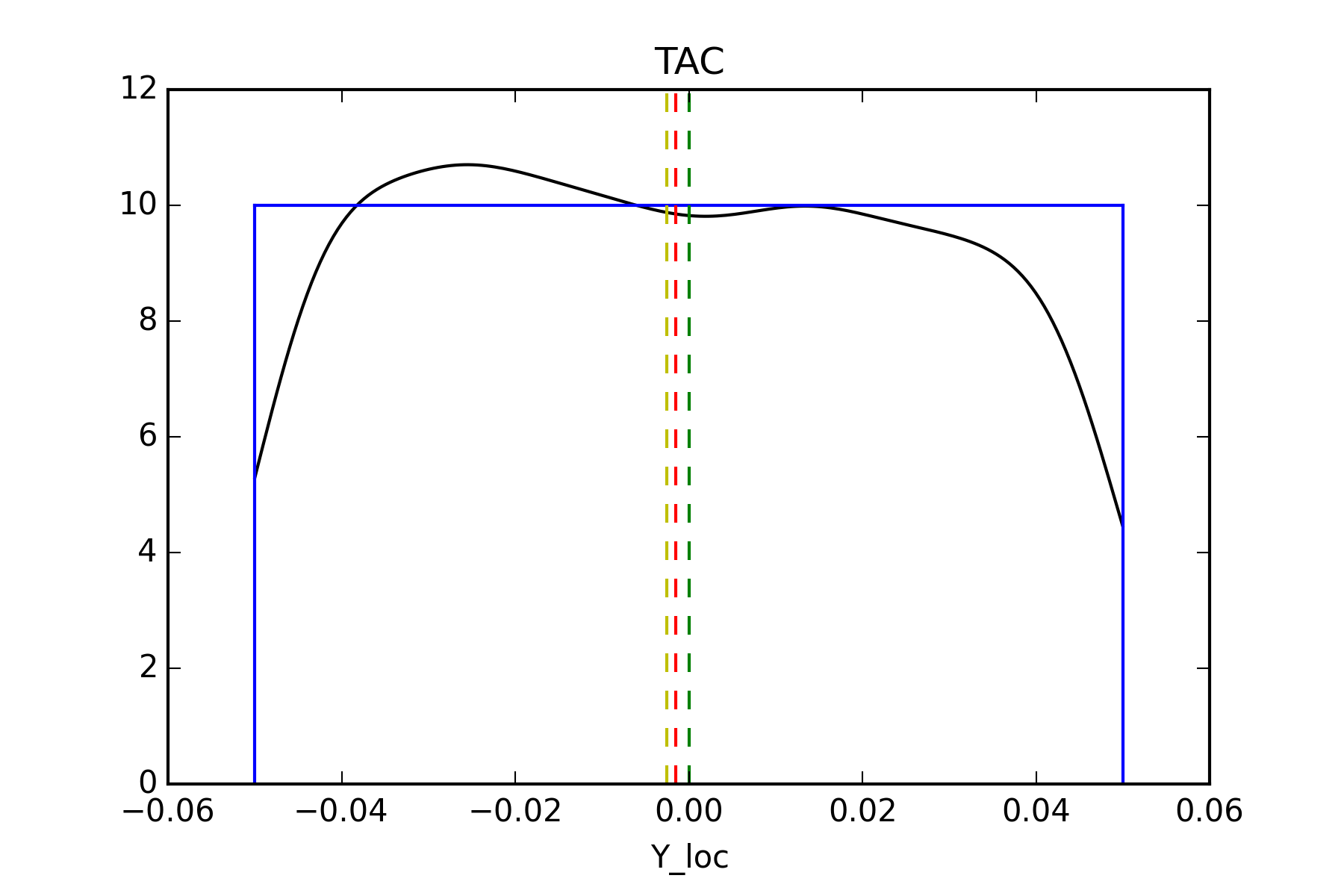}
\includegraphics[width=0.4\textwidth]{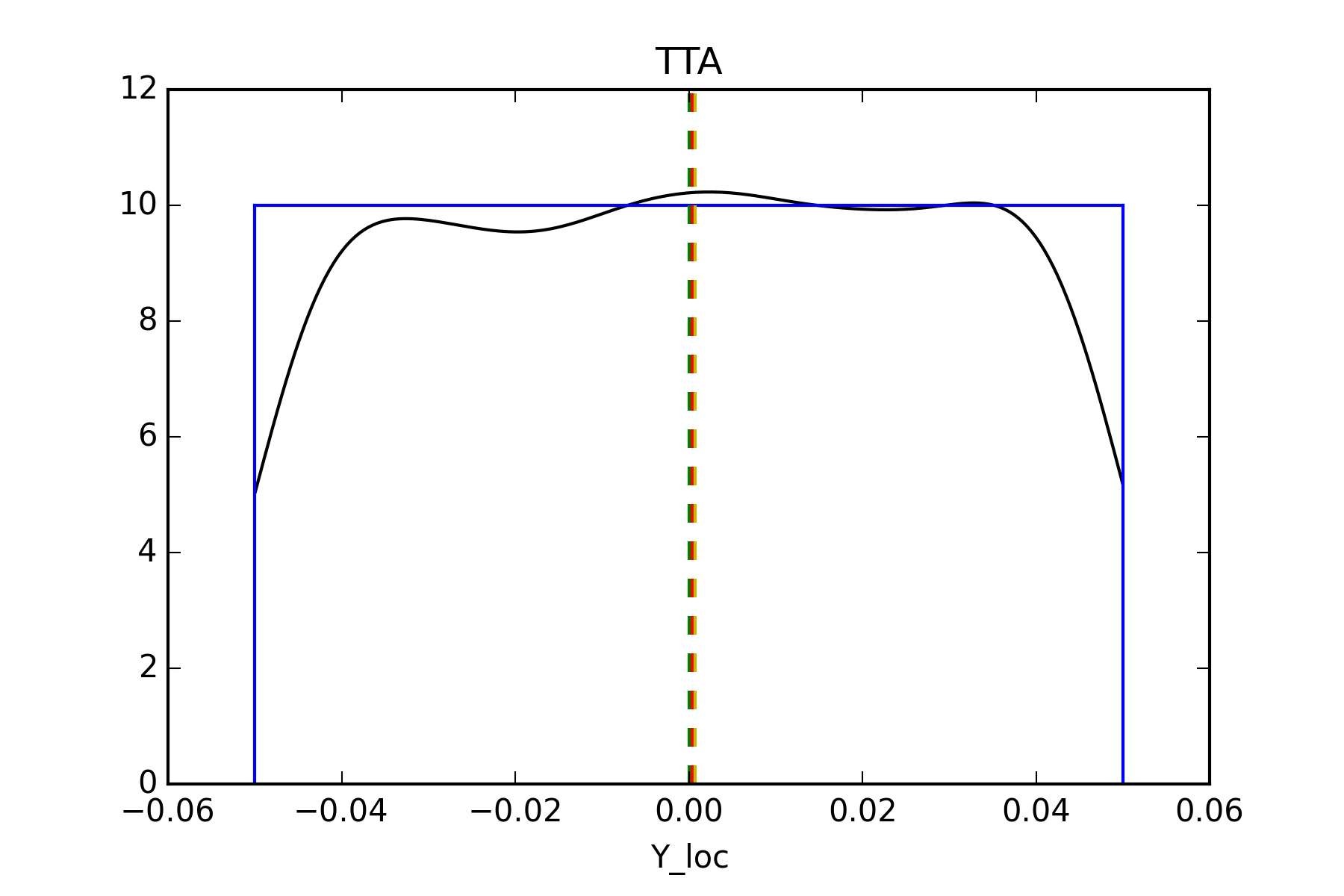}}\\
\resizebox{!}{0.15\textheight}{
\includegraphics[width=0.4\textwidth]{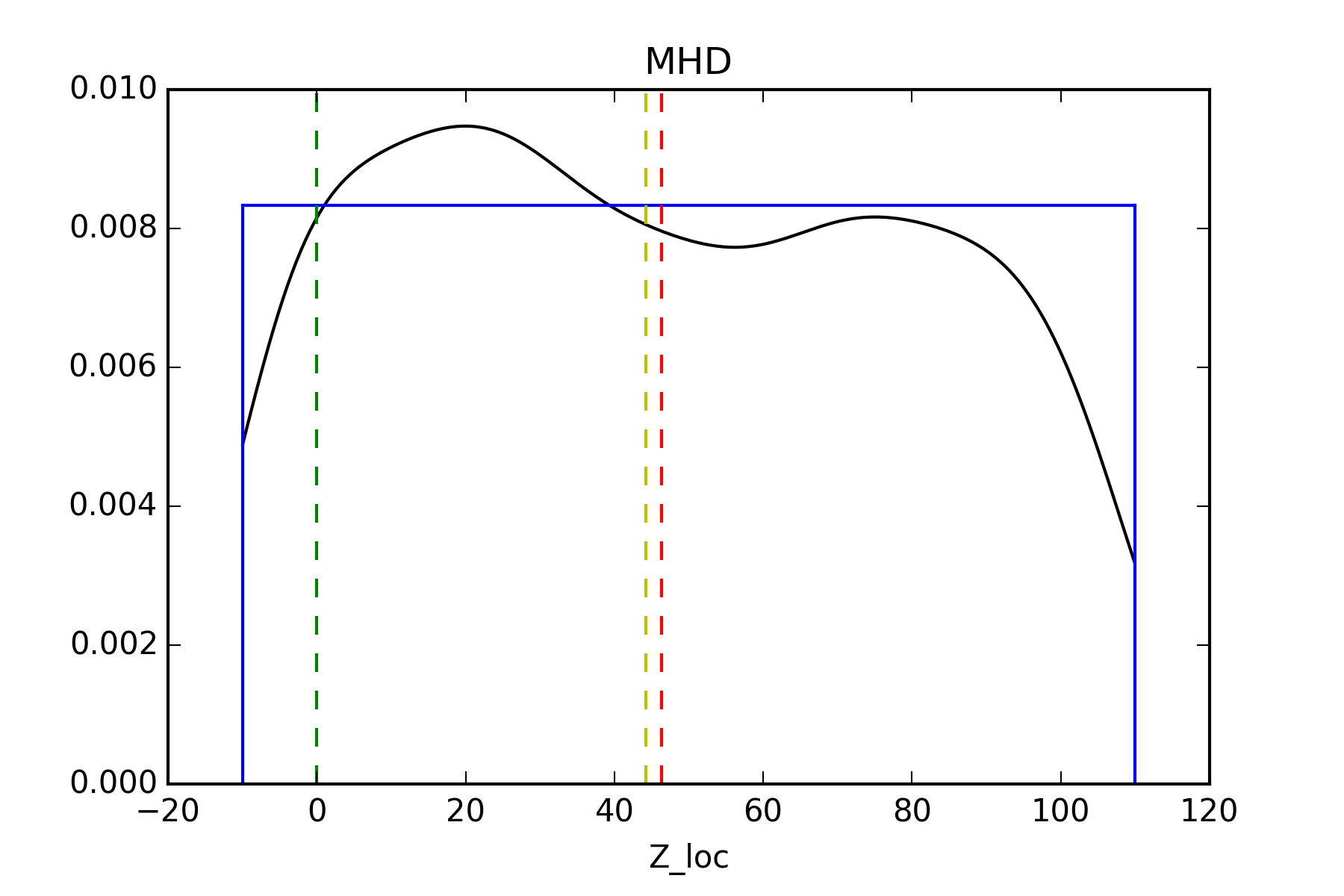}
\includegraphics[width=0.4\textwidth]{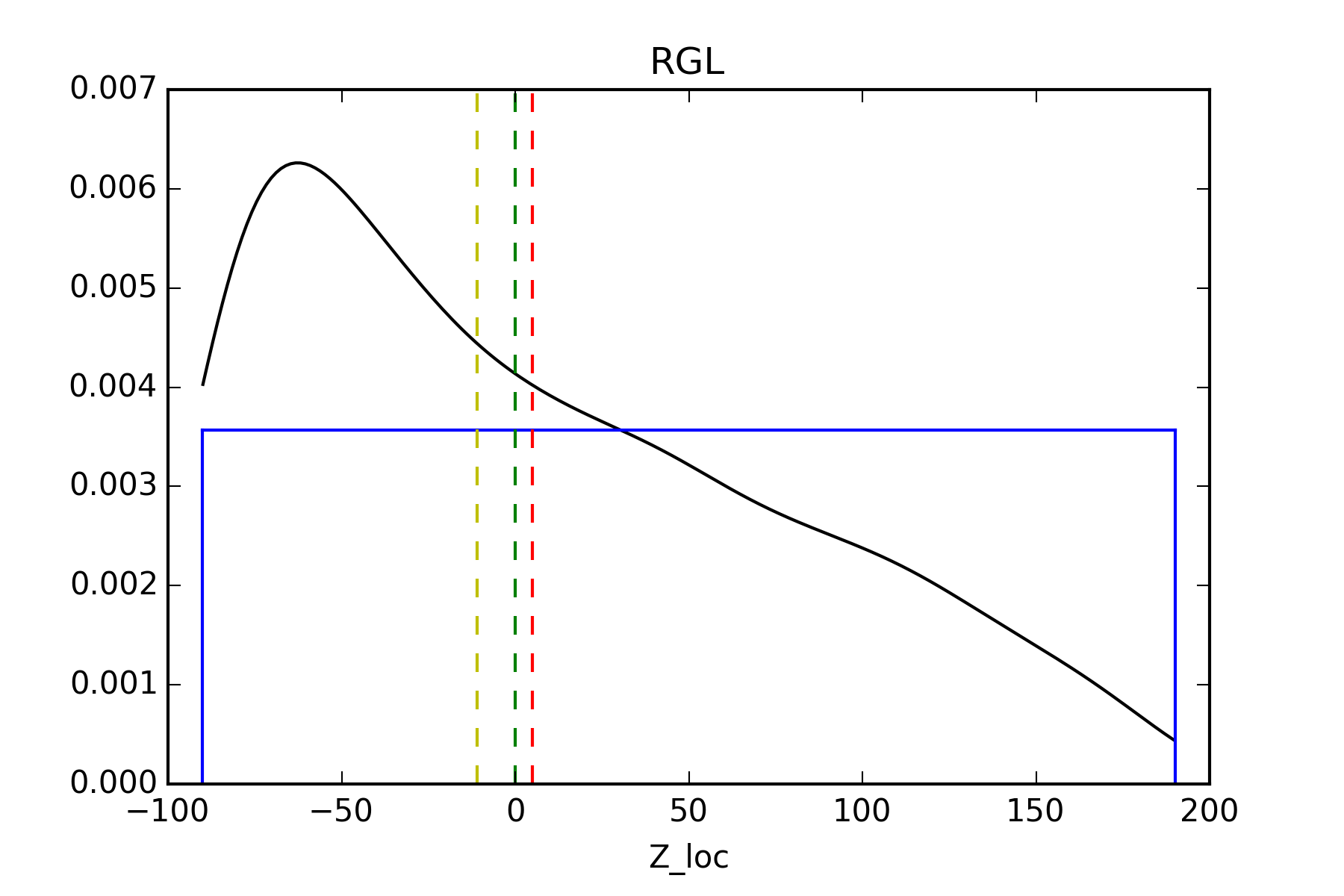}}\\
\resizebox{!}{0.15\textheight}{
\includegraphics[width=0.4\textwidth]{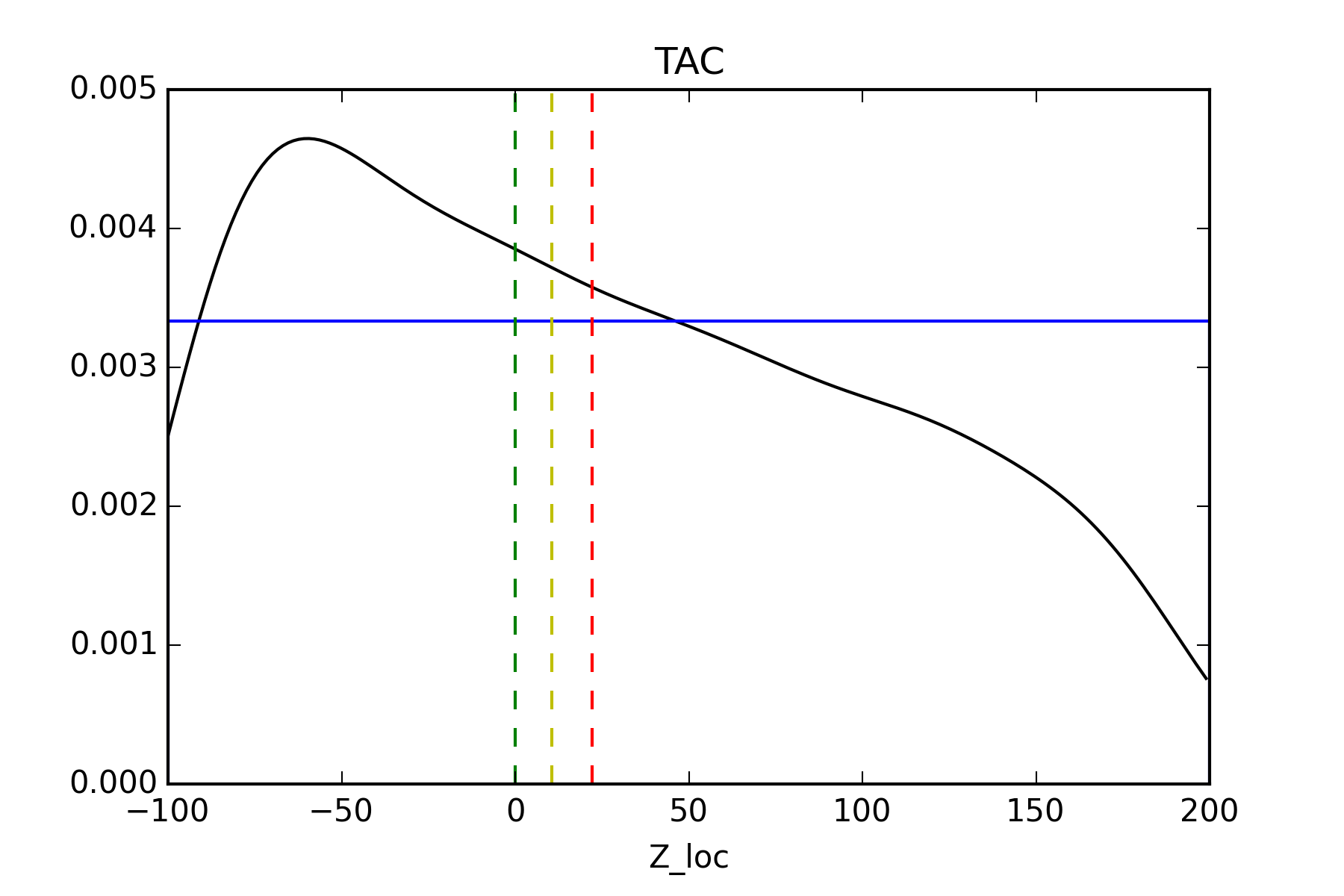}
\includegraphics[width=0.4\textwidth]{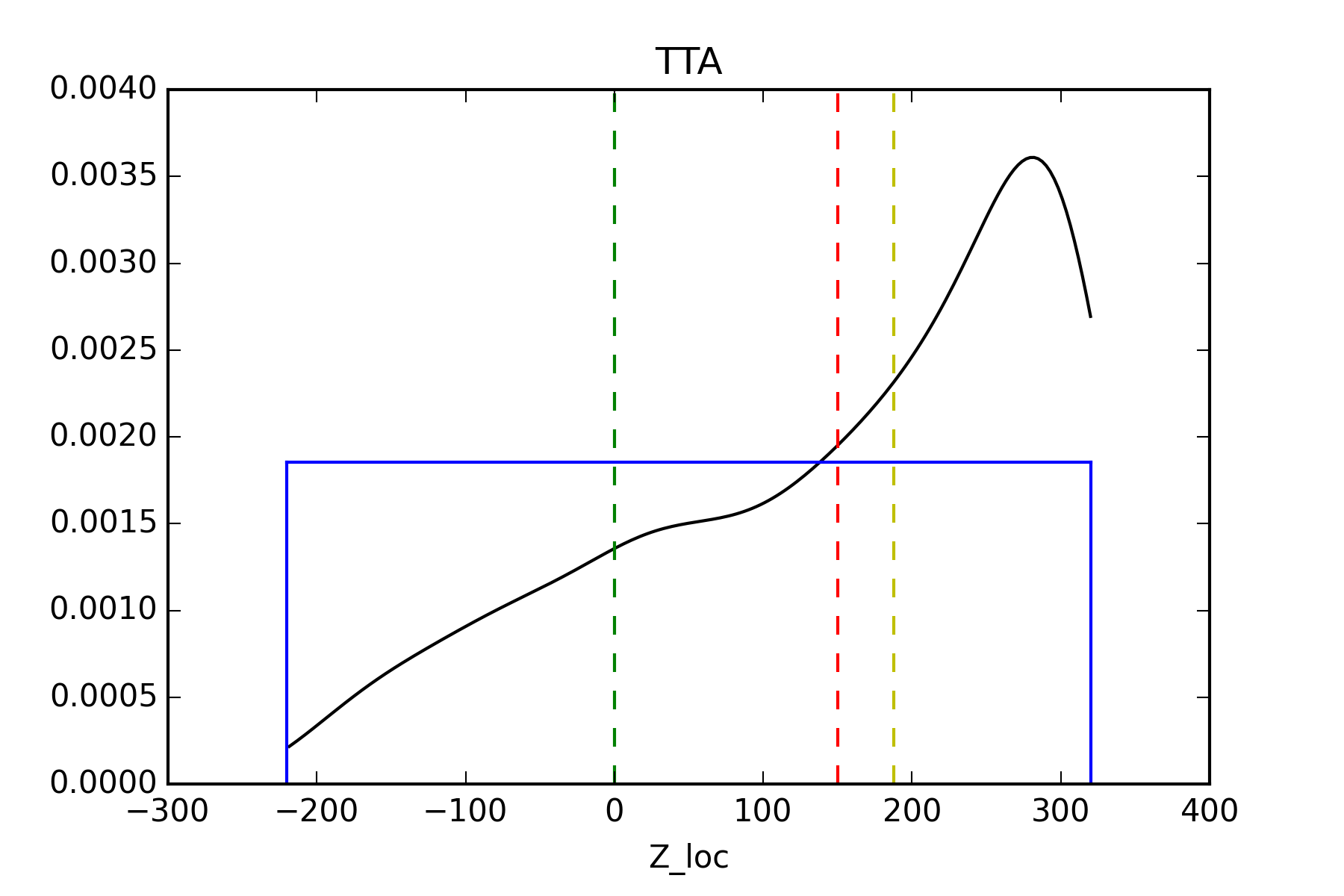}}\\
\caption{Posterior marginal distributions for the site-specific parameters (black). Prior distributions are shown (blue), posterior mean (red) and median (yellow) and the default fixed parameter value (green).}
\label{fig:postmarspec}
\end{figure}

Figures \ref{fig:postmarinv} and \ref{fig:postmarspec} show the posterior marginal distributions for the NAME input parameters.  It is evident that the free tropospheric turbulence and site release height are the two parameters that can be most readily constrained by the data.  Free tropospheric turbulence appears to show the most marked difference from the default value usually used in NAME, with the current default value being completely inconsistent with the posterior distribution for this parameter.  This is consistent with the model having much greater turbulence than is currently accounted for.  For the release height parameters, there is some variability with the constraint on the parameters across sites. TTA shows the biggest discrepancy, whilst also being the site most readily constrained by the data. Conversely for MHD the full prior appears to be returned.  The discrepancy for TTA suggests the model should be releasing particles approximately 150m higher than it currently is.

\clearpage

\begin{figure}[!ht]
\includegraphics[width=\textwidth]{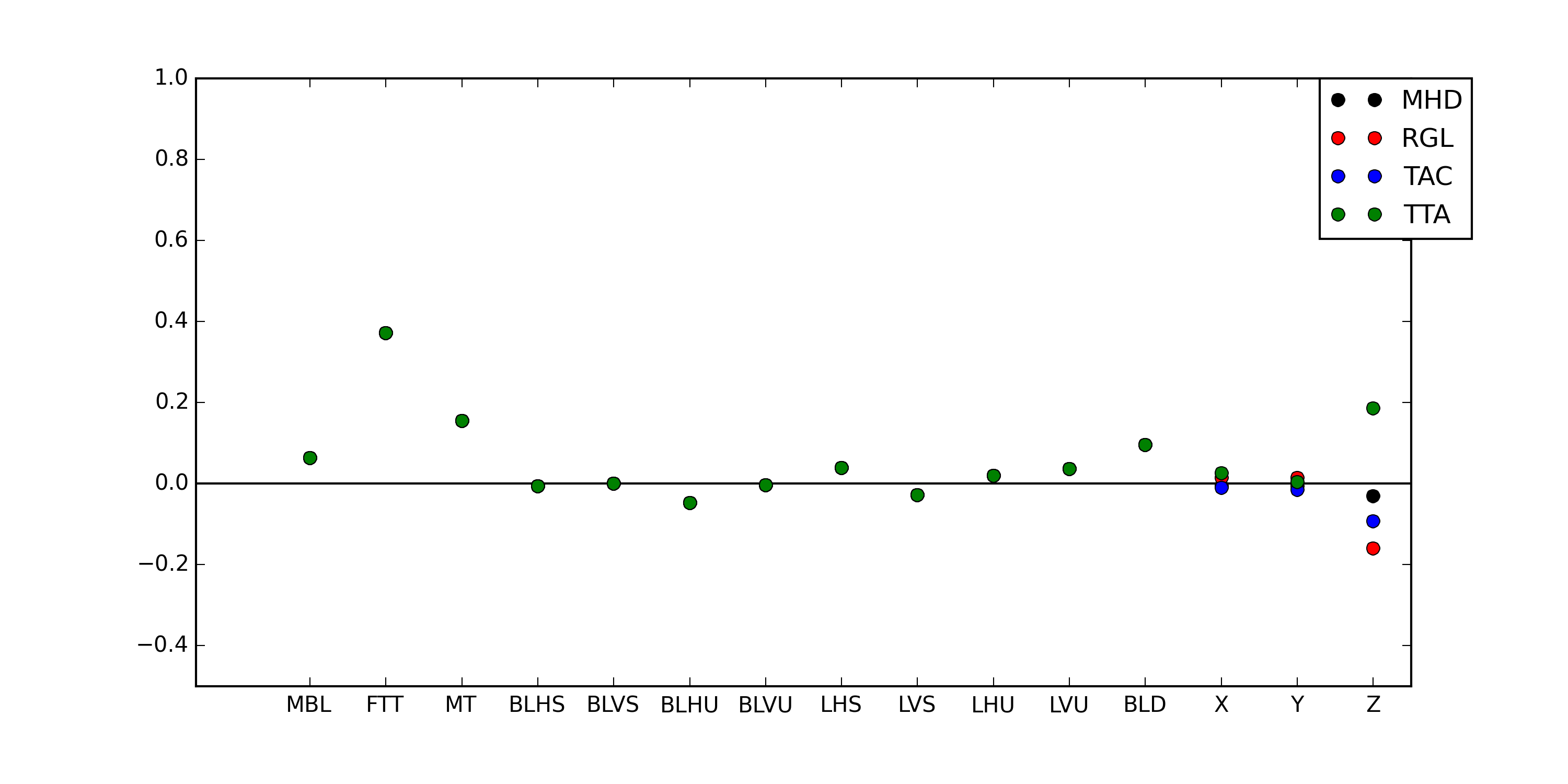}
\caption{Difference between posterior and prior means (posterior minus prior) scaled by the width of the uniform prior interval for each of the 15 parameters.  The site-specific parameters are plotted separately for each site.}
\label{fig:priminpost}
\end{figure}

\begin{figure}[!ht]
\centering
\resizebox{\textwidth}{!}{
\includegraphics[width=0.5\textwidth]{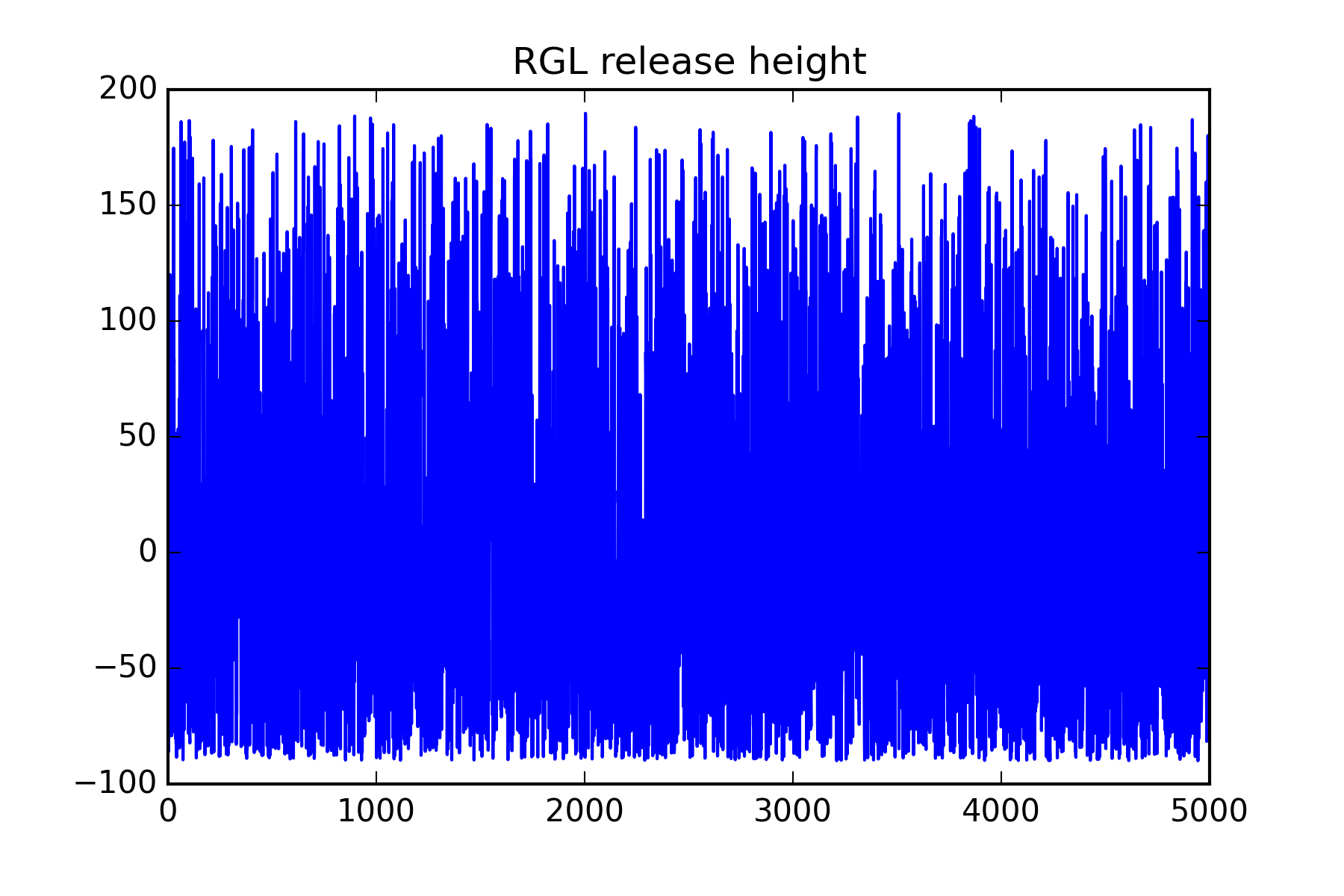}
\includegraphics[width=0.5\textwidth]{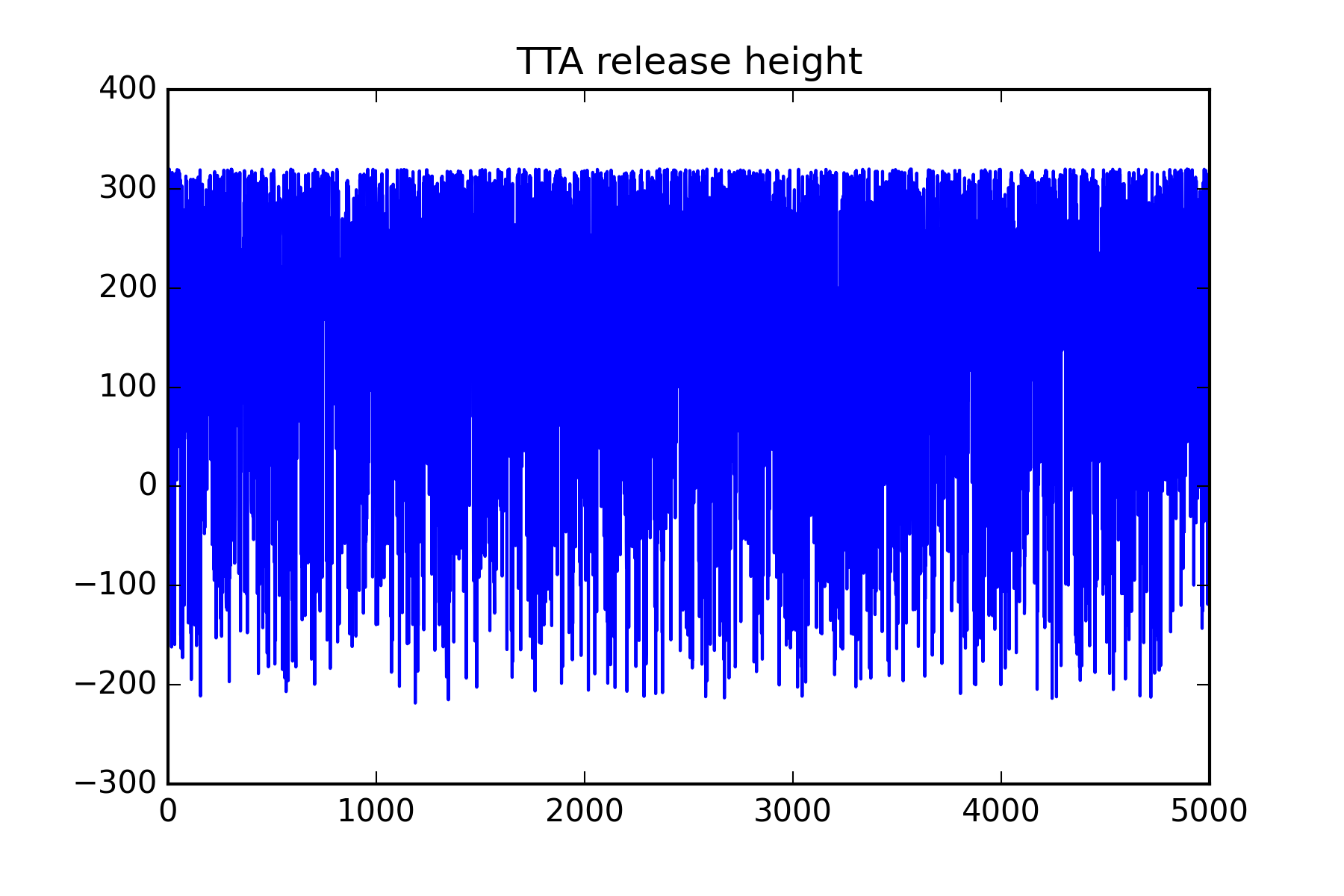}}
\caption{Thinned post-burn-in posterior traceplots for the release height parameter at Ridge Hill (left) and Angus (right).}
\label{fig:posttrace}
\end{figure}

These differences can be more easily seen in Figure \ref{fig:priminpost}, which shows the difference between prior mean and posterior mean scaled by the width of the prior interval.  From this it is again evident that the free tropospheric turbulence parameter differs most from its prior mean value.  This is followed by the release height for Angus (TTA) and the unresolved motion parameter.  The tropospheric turbulence parameter has considerably higher posterior value than both the prior mean value and the current default in NAME.  There is some suggestion from the traceplot in Figure \ref{fig:posttrace} that an even higher value may be plausible above and beyond the upper limit for release height at Angus, whilst the same plot for free tropospheric turbulence indicates a similar idea (not shown).

The majority of the parameters affecting the boundary layer dynamics are relatively unconstrained by the data meaning there is little information in the data to constrain these parameters and/or they have little additional effect on the estimate of $H$ once the other parameters have been accounted for.  The posterior marginals obtained for these parameters fully span the prior range of values showing there is little information in the observations to constrain these parameters.

The estimated UK annual total of methane emissions shows a small increase in magnitude when uncertainty in $H$ is incorporated into the inversion.  The striking change is in the right-hand upper estimate of the 90\% credible interval.  This suggests that the estimate of methane emissions for the UK could potentially be significantly higher than that previously considered.  By incorporating real uncertainties in the simulator used to map the relationship between emissions and the observations, there is a parallel increase in the uncertainty in the estimate of emissions.  This corresponds to a large increase in the upper end of the plausible range for the total emissions.

\begin{figure}[!ht]
\centering
\includegraphics[height=0.3\textheight]{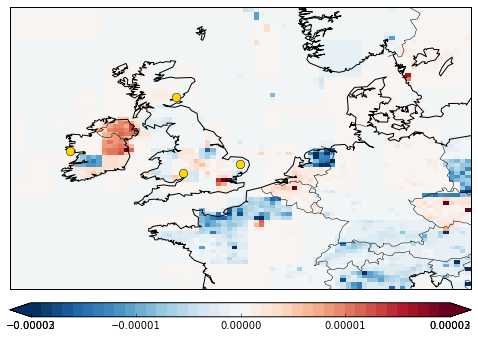}
\caption{Map showing absolute change in posterior mean, i.e. difference in regional posterior means from new method minus regional posterior means assuming fixed $H$, multiplied by prior emissions (from EDGAR and additional wetlands).}
\label{fig:abschange}
\end{figure}

\begin{figure}[!ht]
\centering
\includegraphics[height=0.3\textheight]{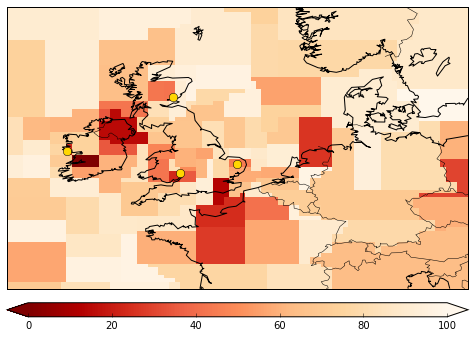}
\caption{Percentage overlap of flux 90\% credible intervals between the inversions with and without uncertainty in $H$.}
\label{fig:CIoverlap}
\end{figure}

\begin{figure}[!ht]
\centering
\includegraphics[height=0.2\textheight]{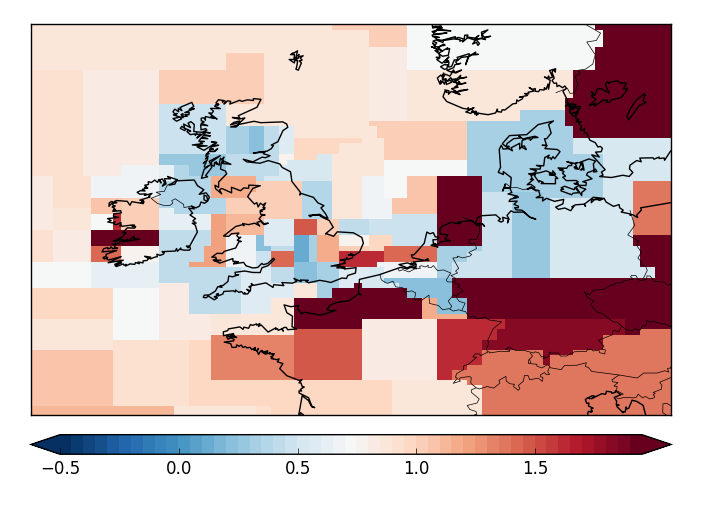}
\includegraphics[height=0.2\textheight]{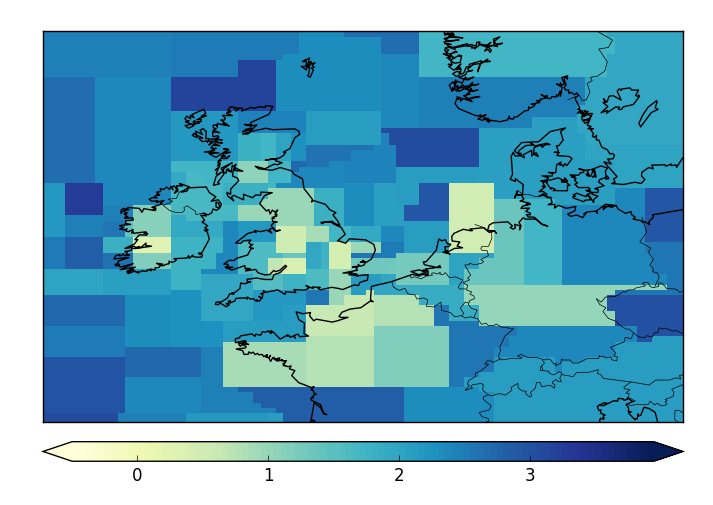} \\
\includegraphics[height=0.2\textheight]{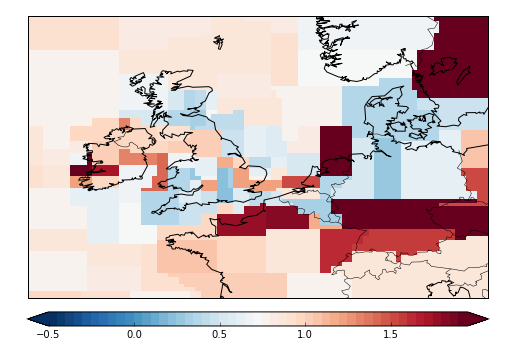}
\includegraphics[height=0.2\textheight]{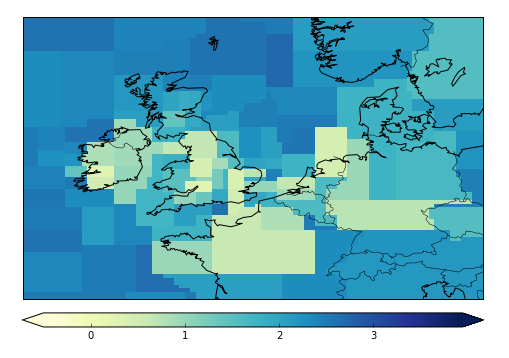}
\caption{Posterior estimate of scaling of prior fluxes estimate for fixed $H$ analysis (top) vs emulator analysis (bottom).  The left-hand plots show posterior means of the scaling whilst the right-hand plot shows the width of the 90\% credible interval scaled by the posterior mean for each region.}
\label{fig:fluxes}
\end{figure}

Figure \ref{fig:abschange} shows the change in posterior means between the two inversions, that is the mean of the fixed $H$ inversion subtracted from the mean of the case with uncertainty in $H$.  This map also incorporates the prior flux estimates from a bottom up approach.  In a UK context, it is clear that the regions representing Northern Ireland and London show the greatest (positive) discrepancy between the two inversions.  Southwest Ireland also seems to show a large negative difference.  To allow for the overall homogeneity in uncertainty in estimates across regions, Figure \ref{fig:CIoverlap} looks at the percentage overlap in 90\% credible intervals.  A value of zero suggests the intervals are completely distinct, or there has been a significant overall change in estimate as a result of incorporating uncertainty in $H$; a value of a hundred suggests that this region has not changed at all.  What is particularly noticeable from this plot is that significant changes are evident in the small regions around the monitoring locations that are not as prevalent in Figure \ref{fig:fluxes}.  The regions closest to the sites will have the smallest uncertainties and hence whilst the absolute difference in flux estimates may be quite small, relatively speaking there is comparatively large changes in these regions.

To ensure that good mixing had been achieved across parameters, acceptance ratios were checked for all estimated parameters, as were traceplots of parameter values (see Figure \ref{fig:posttrace} for examples).  There was no evidence of poor mixing and the tuned proposal variances, which differed from their initial values, appeared to allow good coverage of parameter space as well as ensuring acceptance ratios were adequate.  

\clearpage

\section{Conclusions}
The method developed here directly accounts for uncertainty in the simulator's parameterisations, and therefore the estimates of the magnitude and corresponding uncertainties in the fluxes will be more realistic than those calculated when assuming the $H$ matrix is fixed and not subject to uncertainties.

There are several principal conclusions that can be taken from this study.  It is important to note, however, that these conclusions are purely based on the user-defined specifications of NAME chosen for this particular analysis, such as the turbulence scheme employed; measurement/footprint resolution; month and year; meteorology resolution etc.  Further analyses will be run to observe the consistency (or otherwise) of these results across multiple model specifications.  

The first is the effect of changes in some of the parameters on the output of NAME. What is perhaps surprising is that many of the parameters had little effect on the output, in particular those relating to the boundary layer schemes in NAME.  The nature of the turbulence schemes in the simulator ensure there is no steep gradient between atmospheric levels. The additional limit parameters were added to the parameter study to attempt to test the dependencies between these two schemes.  However, there were multiple dependencies based on the Lagrangian timescales, the velocity variances and the diffusivities.  Only the dependencies in diffusivities were varied and hence there is still likely some remaining dependence between the two schemes.  

It is therefore perhaps not too surprising that the boundary layer turbulence parameters are not readily being constrained by the data beyond the prior ranges given.  If the free tropospheric turbulence (FTT) value is being increased, then any changes in the boundary layer (BLT) may not be implemented within the code due to its dependency on the free troposphere turbulence value. For example, if the inputted value of FTT was 0.2 and a given BLT was 0.1, the value that NAME would use for that BLT may be 0.1 if the two schemes were deemed sufficiently similar.  Similarly, if FTT was 0.8 and BLT was 0.05, then NAME would most likely increase the value of BLT to ensure the step between the two schemes wasn?t too great.  In terms of the emulator, the values of the two parameters it sees are still 0.8 and 0.05, but the estimate of the effect of changing FTT may be increased as it will be having an effect on both schemes in this instance.  This will be reflected in the elements of the $B$ matrix relating to the relationship between input and output spaces.  Hence, the dominance of FTT over the other parameters could expected. If the upper limit of FTT was cut off at a lower value, or more weight was given to the lower part of the range, BLT parameters could potentially show a greater importance. 

Secondly, the two parameters that appear to have the largest effect and can also be best constrained by the observations, seem to be different from the values currently used as defaults in NAME. The most severe difference observed was for the free tropospheric turbulence parameter, for which the posterior credible interval was completely distinct of the parameter value currently used.  To a less significant degree, the release height also differs from that currently used for TTA and to a lesser degree RGL.  The value for TTA suggests that the particles should be released significantly higher in the computational domain than they currently are being.  The estimate of the height difference estimated from the atmospheric measurements corresponds well with the difference in value between the modelled ground height for the grid cell that contains the site and the true ground height at the observation site.  

Clearly the aim of the model developers is to produce the most physically accurate model possible for the processes governing atmospheric transport. It is important to note that the estimates of the parameters gained from the statistical model may not be representing specific physical processes but merely the combination of parameters that best match the measured data. There is some evidence to suggest that when comparing with observations, a better model fit can generally be achieved agreement by adding in some extra spread through the turbulence (H.N. Webster, \emph{pers. comm.}). There is no physical reason for this, but more likely due to the fact that transport errors, (e.g., due to errors in the meteorology), may put the material in the wrong place so extra spread may spread material to the observation.

Interestingly, the absolute change in flux estimate does not tell the full story.  The regions closest to the sites, where most information will be available from the data, may not change particularly in absolute value, but when accounting for the uncertainty in these regions, the overlap of credible interval plot (Figure \ref{fig:CIoverlap}) shows many of the regions close to the observation sites have changed quite significantly.

For some of the parameters, there was evidence from the traceplots and posterior marginal distributions that the  parameters were pushing up against the upper boundaries of the prior (Figure \ref{fig:posttrace}).  If there are physical reasons for not allowing the parameters to go above this value, then evidently greater uncertainty in these parameters in particular should be studied further.  However, this may suggest areas of further discussion as to whether the upper bounds could be realistically extended.  In addition, some of the posterior marginal distributions appear to show a tendency for the very extreme values at the upper and lower end of the prior ranges not to be visited by the MCMC chain.  A change of proposal distribution to improve the number of proposals into the extremes may be warranted \citep{andrieu08}, although the affect of this on the results should be minimal.

In general, the values of the upper end of the range for free tropospheric turbulence parameter are expected to be high turbulence in pockets rather than a typical average value.  As such, it is probably unrealistic to allow this parameter to be greater than the range currently studied.  Additionally, a uniform prior distribution may not be particularly valid for this parameter.  Using a prior distribution with more weight in the lower end of the range would be worth testing to see if the importance of the boundary layer turbulence parameters increases to compensate for this or whether the information in the data is sufficiently strong to counteract this.

The methodology developed here has two main benefits over assuming the simulator produces a single fixed output.  Firstly, as is clear from the marginal posterior distributions, the values for at least two of the input parameters were significantly different from those currently being used in NAME. The method therefore acts as a `verification' of the computer code and its inputs currently being used \citep{rougier09}. Further analyses will confirm whether this is consistently the case across other months/years and will allow direct comparison with previous estimates of methane from the same sites conducted by \citet{ganesan15} but assuming $H$ is fixed. As mentioned previously, it is necessary to determine further the validity of these parameter estimates to understand why the model is pushing the value of these parameters higher.  If this is a symptom of, for example, poorly defined meteorology, then improving the underlying cause is evidently the necessary course of action to prevent systematic biases being introduced into our overall statistical inversion framework.  

The fact that the modelled ground height for sites such as Angus (TTA) is approximately 200m below the true height correlates well with the estimate for this parameter from independent data (that from the inversion).  Although the above methodology highlights the specific discrepancy between release heights for Angus, it represents an overall issue with the resolution of topography files that is currently most commonly implemented by users of the simulator.  Further study to observe what effect changing the resolution of the meteorology (and hence topography) files has on the results would also be of interest.  Whilst it would be possible to release particles from 200m higher than currently by changing the input parameters, it would clearly be advantageous to attempt to solve the underlying problem of the topography files.  Either using the height specifically at the observation station as the ground height rather than an interpolated value, or an increase in the resolution could help reduce the discrepancies currently introduced into the model.  Modelling other sites with even more complex topography will be potentially introducing even greater systematic errors if this is not improved. This is not a direct issue of the simulator itself, but a problem with the files that are fed into it.

\subsection{Further developments}
There are several ways to take this work further, both with NAME and applying the methodology to other simulators.  Firstly, we will apply this method to additional months and years to check the consistency of this approach over the seasons and with other observations sites.  Previous problems observed when adding additional sites to the inversion may be avoided or the magnitude of their effects be reduced when uncertainties in NAME have been accounted for.

We can also incorporate this emulation framework into that developed by \citet{lunt16}, where the dimension of the $\mathbf{x}$ vector is also an unknown.  The reversible jump algorithm is used in this case to select the number, positioning and size of each of the regions.

Secondly, there are many other ways of calculating the statistical relationship between the (dimension-reduced) simulator output and the input parameters.  Exploring different techniques to produce the best statistical prediction is an area of interest.  One possible change would be looking at alternative procedures for reducing the dimension of the input parameters. Here we have used AIC as a means for deciding whether a parameter should be included or not in the estimation of each singular value.  In cases where the parameter was not selected, the effect of that parameter was fixed at zero (such as in Table \ref{tab:AICbin}).  Methods that scale the effect of parameters rather than removing them entirely may offer a better method for predicting variation in the singular values.

One such method is (gKDR) developed by \citet{fukumizu14} and used in GP emulation by \citet{liu16}.  In the case of gKDR, all input parameters would affect each of the singular values, but would be scaled up or down according to their importance.  This would have a similar effect to AIC but would maintain all parameters and should reduce the problem of too much noise being introduced into the inversion as was previously observed when no parameter selection was included.  Other selection methods could also be used.  If we were to continue using SVD, it would also be interesting to look at the effect of using a threshold value to cut off the singular values at only the most important ones, hence further reducing the dimension of the problem.  

Each simulator will require a slightly different approach, particularly in the dimension reduction and statistical model setup stages, depending on the nature of the problem at hand.  For example, working on the Orbiting Carbon Observatory-2 (OCO-2) satellite forward model requires a very large vector of wavelengths to be emulated.  SVD would not be appropriate in this case, but other alternative decomposition techniques may work better.

\section*{Acknowledgements}
We wish to thank David Thomson and Helen Webster for their expertise utilised in the expert elicitation, as well as helpful comments on this manuscript. We are also greatful to Helen Dacre and Natalie Harvey for originally conducting this elicitation.  We would also like to thank all the site operators of the DECC network for their tireless work and dedication to providing high quality and reliable data. Matt Rigby is funded by a NERC advanced research fellowship NE/I021365/1.
\bibliographystyle{plainnat}
\bibliography{sample}

%

\end{document}